\documentclass[]{article}
\usepackage[margin=1in]{geometry}

\usepackage[utf8]{inputenc}
\usepackage[pdftex, pdftitle={Article}, pdfauthor={Author}]{hyperref} 
\usepackage[parfill]{parskip}
\usepackage{float} 
\usepackage{array}
\usepackage{multirow}

\newcolumntype{C}{>{\centering\arraybackslash}p{1 cm}}
\newcolumntype{V}{>{\centering\arraybackslash}p{2 cm}}
\newcolumntype{B}{>{\centering\arraybackslash}p{2.8 cm}}
\usepackage{chemformula}
\usepackage{graphicx}
\usepackage{amsmath,amssymb,amsthm}
\usepackage[
backend=biber,
style=nature,
citestyle=numeric-comp,
sorting=none,
autocite=superscript,
]{biblatex}

\usepackage[utf8]{inputenc}
\usepackage{setspace} 
\usepackage{authblk}

\usepackage{svg}
\newcounter{extendedfigure}
\newenvironment{extendedfigure}
  {\stepcounter{extendedfigure}\begin{figure}}
  {\end{figure}}
  
\usepackage{titlesec}
\titleformat{\part}
  {\normalfont\Huge\bfseries\centering} 
  {\partname\ \thepart}{20pt}{\Large}

\title{\fontsize{15}{16}\selectfont\textbf{Superfluid Stiffness and Flat-Band Superconductivity in Magic-Angle Graphene Probed by cQED}}
\date{}
\author[1,2,$\dagger$]{Miuko~Tanaka}
\author[1,$\dagger$*]{Joel~\^I-j.~Wang}
\author[3]{Thao~H.~Dinh}
\author[1]{Daniel~Rodan-Legrain}
\author[1,4]{Sameia~Zaman}
\author[1]{Max~Hays}
\author[1,4,8]{Bharath~Kannan}
\author[1,4]{Aziza~Almanakly}
\author[5]{David~K.~Kim}
\author[5]{Bethany~M.~Niedzielski}
\author[1,5]{Kyle~Serniak}
\author[5]{Mollie~E.~Schwartz}
\author[6]{Kenji~Watanabe}
\author[7]{Takashi~Taniguchi}
\author[1]{Jeffrey~A.~Grover}
\author[1,4]{Terry~P.~Orlando}
\author[1,8]{Simon~Gustavsson}
\author[3*]{Pablo~Jarillo-Herrero}
\author[1,3,4*]{William~D.~Oliver}

\affil[1]{Research Laboratory of Electronics, Massachusetts Institute of Technology, Cambridge, MA 02139, USA}
\affil[2]{Institute for Solid State Physics, The University of Tokyo, 5-1-5 Kashiwanoha, Kashiwa, Chiba, Japan}
\affil[3]{Department of Physics, Massachusetts Institute of Technology, Cambridge, MA 02139, USA}
\affil[4]{Department of Electrical Engineering and Computer Science, Massachusetts Institute of Technology, Cambridge, MA 02139, USA}
\affil[5]{Lincoln Laboratory, Massachusetts Institute of Technology, 244 Wood Street, Lexington, MA 02421, USA}
\affil[6]{Research Center for Electronic and Optical Materials, National Institute for Materials Science, 1-1 Namiki, Tsukuba 305-0044, Japan}
\affil[7]{Research Center for Materials Nanoarchitectonics, National Institute for Materials Science,  1-1 Namiki, Tsukuba 305-0044, Japan}
\affil[8]{Present address: Atlantic Quantum, Cambridge, MA 02139, USA}

\begin{document}

\maketitle
\normalsize{$^\dagger$These authors contributed equally to this work.}\\
\normalsize{$^\ast$To whom correspondence should be addressed:\textcolor{blue}{joelwang@mit.edu, pjarillo@mit.edu,} and \textcolor{blue}{william.oliver@mit.edu}}
\pagebreak
\section*{Abstract}
The physics of superconductivity in magic-angle twisted bilayer graphene (MATBG) is a topic of keen interest in moir\'e systems research, and it may provide insight into the pairing mechanism of other strongly correlated materials such as high-$T_{\mathrm{c}}$ superconductors. 
Here, we use DC-transport and microwave circuit quantum electrodynamics (cQED) to measure directly the superfluid stiffness of superconducting MATBG via its kinetic inductance. 
We find the superfluid stiffness to be much larger than expected from conventional Fermi liquid theory; rather, it is comparable to theoretical predictions involving quantum geometric effects that are dominant at the magic angle~\autocite{torma2022superconductivity}. 

The temperature dependence of the superfluid stiffness follows a power-law, which contraindicates an isotropic BCS model;
instead, the extracted power-law exponents indicate an anisotropic superconducting gap, whether interpreted within the Fermi liquid framework or by considering quantum geometry of flat-band superconductivity.
Moreover, a quadratic dependence of the superfluid stiffness on both DC and microwave current is observed, which is consistent with Ginzburg-Landau theory.
Taken together, our findings indicate that MATBG is an unconventional superconductor with an anisotropic gap and strongly suggest a connection between quantum geometry, superfluid stiffness, and unconventional superconductivity in MATBG.
The combined DC-microwave measurement platform used here is applicable to the investigation of other atomically thin superconductors. 

\section*{Main}

\subsection*{Introduction}
Stacking two graphene sheets with a finite twist angle between individual crystallographic axes forms a moir\'e superlattice. 
In the special case known as magic-angle twisted bilayer graphene (MATBG), a twist angle of approximately 1.05° (the first magic angle) results in flat energy bands, which facilitate strong electron-electron interactions.
Multiple phases of matter have been observed in MATBG, including correlated insulators, superconductors, strange-metal phases, and topological insulating states, all accessible via an applied gate-voltage that controls the carrier density~\autocite{cao2018correlated, Cao2018, cao2020strange, balents2020superconductivity, andrei2020graphene,ma2020moire,}.

Of particular interest is the superconductivity observed in MATBG 
~\autocite{park2021tunable,hao2021electric,park2022robust, zhang2022promotion, burg2022emergence}, as it shares a notable resemblance to other unconventional, interaction-driven superconductors, including cuprates and heavy-fermion superconductors~\autocite{Cao2018, oh2021evidence, cao2021nematicity, balents2020superconductivity, andrei2020graphene}. 
Observing superconductivity in the flat-band regime suggests that it is driven by the superfluid stiffness determined not only by the band dispersion, but also by the band geometry in phase space~\autocite{peotta2015superfluidity, hu2019geometric, julku2020superfluid, xie2020topology, torma2022superconductivity, hofmann2022heuristic}. 
Understanding superconductivity in MATBG could therefore provide fundamental insights into the mechanisms that underlie unconventional superconductivity in other materials.

However, to date, there have been relatively few experimental investigations of the unconventional pairing in superconducting moir\'e systems.
Many of the conventional approaches used to study the gap structure of bulk superconductors, such as the magnetic penetration depth, Meissner effect, thermal transport, and inelastic neutron scattering, are difficult to apply to van der Waals (vdW) heterostructures made by mechanical exfoliation.
This is due in part to the small size of typical 2D samples -- atomically thin with a few-micron-square area and an inhomogeneous twist angle -- and their generally ultra-low carrier density\autocite{andrei2021marvels}.

The electromagnetic properties of a superconductor can be described by the constitutive relation (in the London gauge)
\begin{equation} \label{eqn:london}
    \boldsymbol{j}=-D_{\mathrm{s}}\boldsymbol{A},
\end{equation} 
where $D_{\mathrm{s}}$ is the superfluid stiffness, \textbf{j} is the 2D current density (current per unit width), and \textbf{A} is the vector potential\autocite{tinkham2004introduction}. 
The inverse of the superfluid density is the kinetic inductance $L_{K} = 1/D_{\mathrm{s}}$. Therefore, one can directly access $D_{\mathrm{s}}$ by measuring $L_{K}$ in an AC circuit. 
Within Fermi liquid theory, $D_{\mathrm{s}}$ is directly proportional to the Cooper pair density $n_{\mathrm{s}}$ and inversely proportional to the charge carrier effective mass $m_{\mathrm{eff}}$:
\begin{equation} \label{eqn:lkandns}
    D_{\mathrm{s}}= \frac{2n_{\mathrm{s}}e^2}{m_{\mathrm{eff}}}.
\end{equation} 
 
Moreover, the temperature dependence of $n_{\mathrm{s}}$ is characteristic of the pairing symmetry and gap structure of a superconductor due to its quasiparticle spectrum at finite temperature. Characterizing the kinetic inductance is hence a promising method to study the nature of superconductivity in unconventional bulk superconductors~\autocite{prozorov2006magnetic, hardy2002magnetic} as well as homogeneous superconducting thin films~\autocite{bottcher2024circuit, phan2022detecting, weitzel2023sharpness,}.

In this work, we measure directly the superfluid stiffness via the kinetic inductance $L_{K}$ in MATBG and its dependence on temperature, DC bias current, and microwave power. 
The measured superfluid stiffness is much larger than conventional values estimated by the band dispersion within the Fermi liquid framework; rather, the magnitude of the measured $D_{\mathrm{s}}$ is consistent with theoretical works incorporating quantum geometric contribution that are predicted to dominate at the magic angle~\autocite{torma2022superconductivity}. 
Furthermore, the temperature dependence of the superfluid stiffness exhibits a power-law behavior that contraindicates an isotropic BCS model, as expected for the unconventional superconductivity in MATBG, and consistent with an anisotropic gap under the influence of the quantum-metric in a flat-band superconductor. 
Taken together, our results indicate both an anisotopic gap and a quantum-geometric contribution to the unconventional superconductivity in MATBG.

\subsection*{Device Configuration}
\begin{figure}[H]
    \centering
    \includegraphics[width=0.65\textwidth]{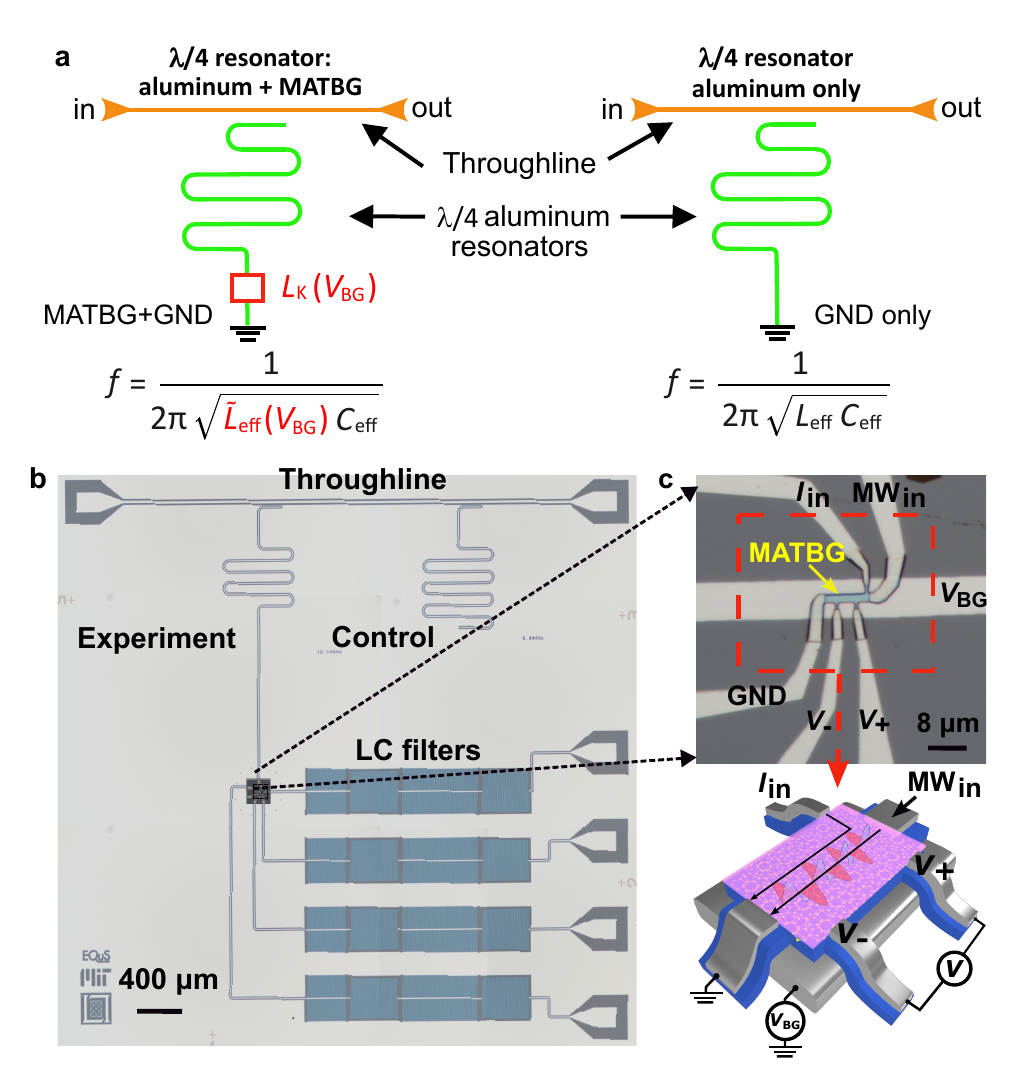}
    \caption{\textbf{Kinetic inductance measurement and device configuration.} 
    \textbf{a,} Schematic of the magic angle twisted bilayer graphene (MATBG) inductance measurement circuit. The aluminum quarter-wave 
 ($\lambda /4$) ``experiment resonator'' terminates to ground through the MATBG sample. The aluminum quarter-wave 
 ($\lambda /4$) ``control resonator'' terminates directly to ground with aluminum.
    Both resonators couple capacitively to the throughline, which is probed by a network analyzer (not shown) to measure the resonance frequencies of the experiment and control resonators. The resonance frequency of the experiment resonator is used to infer the gate-voltage-dependent inductance of the MATBG. For comparison, the control resonator is used to check the inductance characterization technique using an aluminum termination, a well-characterized BCS material (see extended methods).
    \textbf{b,} Optical image of a 5x5 mm$^2$ chip comprising the resonators, throughline, DC bias lines, filters, and the ground plane, patterned from 250-nm thick aluminum on a high-resistivity Si substrate. The DC electrodes are LC-filtered on-chip to reduce bias noise. 
    \textbf{c,} Zoomed-in image and schematic of the MATBG sample on the experiment resonator. The hBN/MATBG/hBN heterostructure is placed on an aluminum backgate to voltage-bias the MATBG and change its carrier density.}
    \label{fig:sample}
\end{figure}

We use a superconducting quarter-wavelength ($\lambda/4$) waveguide resonator to characterize the kinetic inductance of MATBG (Fig.~\ref{fig:sample}).
A $\lambda/4$ resonator terminated directly to ground has a resonance frequency \(f=\frac{1}{2\pi\sqrt{L_{\mathrm{eff}}C_{\mathrm{eff}}}}\), where $L_{\mathrm{eff}}$ and $C_{\mathrm{eff}}$ are the effective inductance and capacitance of the resonator, respectively. 
In contrast, when the resonator is terminated to ground through a MATBG sample, the resonance frequency 
shifts to a new frequency \(f^{'}=\frac{1}{2\pi\sqrt{(\tilde{L}_{\mathrm{eff}}(V_{\mathrm{BG}}))C_{\mathrm{eff}}}}\), due to the appreciable kinetic inductance of the MATBG that is added to the resonator intrinsic inductance, where $\tilde{L}_{\mathrm{eff}}$ includes the additional gate-dependent effective inductance of the MATBG-terminated resonator.
As we describe below, the added inductance of the MATBG sample can be extracted from shifts in the resonance frequency (Fig.~\ref{fig:sample}a).

Fig.~\ref{fig:sample}b shows the superconducting resonators used in this experiment. The resonators are patterned from a 250-nm aluminum film deposited on a high-resistivity silicon substrate. 
A common throughline couples capacitively to both a ``control'' and an ``experiment'' resonator and is used to measure their resonance frequencies. 
The control resonator is used for two purposes: 1) to test the kinetic inductance measurement technique on aluminum, a known isotropic BCS material (see Figure 4 and \ref{fig:cont} for more details); and 2) to assess the impact on the aluminum film due to the MATBG fabrication.
The experiment resonator is terminated with an hBN-MATBG-hBN heterostructure (Fig. \ref{fig:sample}c) 
positioned on an aluminum backgate, to which a voltage is applied to tune the MATBG carrier density. 
In addition, five galvanic contacts -- superconducting at cryogenic temperatures -- connect the MATBG to the microwave resonator, the ground plane, and three DC probe electrodes (Fig. \ref{fig:sample}c).
This design enables us to perform both microwave and DC transport characterization of the MATBG device. 

\subsection*{Measuring the Kinetic Inductance and Critical Temperature of MATBG}

\begin{figure}[H]
    \centering
    \includegraphics[width=1\textwidth]{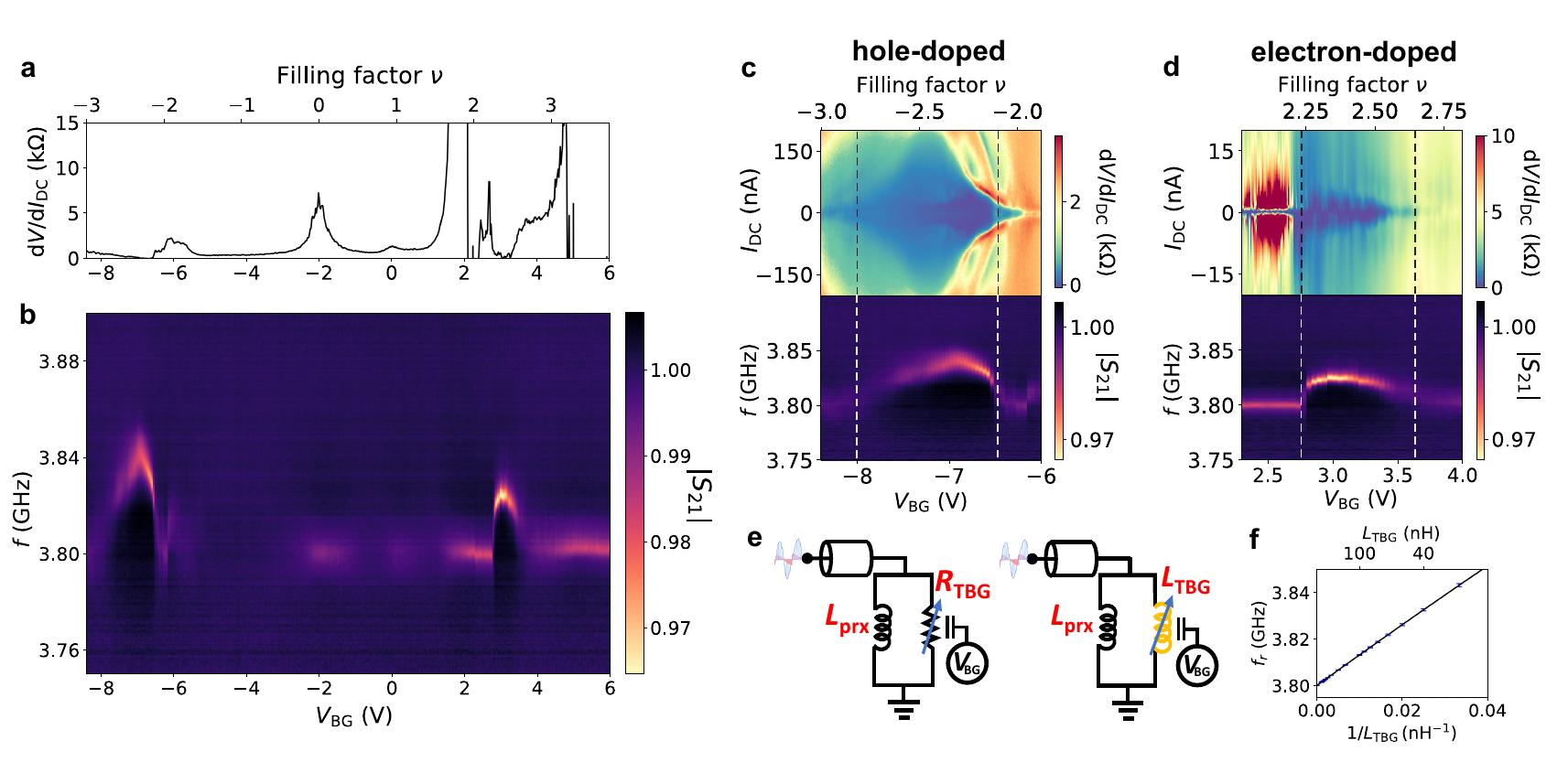}
    \caption{\textbf{Gate-voltage dependent DC and microwave characteristics.} 
    \textbf{a,} Differential resistance $\mathrm{d}V/\mathrm{d}I_{\mathrm{DC}}$ of the MATBG termination at zero bias current as a function of the backgate voltage $V_{\mathrm{BG}}$. Top axis represents the filling factor $\nu$.
    \textbf{b,} Microwave transmission coefficient $|S_{21}|$ versus $V_{\mathrm{BG}}$. The resonant frequency (bright line) shifts within the zero-resistance region in panel (a), near filling factors $\nu=\pm 2$. The resonance remains essentially constant within the high-resistance region. 
    \textbf{c,d, } Differential resistance $\mathrm{d}V/\mathrm{d}I_{\mathrm{DC}}$ (top panel) and frequency shift (bottom panel) as a function of $V_{\mathrm{BG}}$ and $I_{\mathrm{DC}}$ for hole-doped (\textbf{c}) and electron-doped (\textbf{d}) MATBG. Top axis represents the filling factor $\nu$. The superconducting regions lie between the two dashed lines.
    \textbf{e,} Lumped-element model for the MATBG termination of the resonator and the electrode-induced proximity effect when voltage-biased in the highly-resistive regime (left panel) and the superconducting regime (right panel). See main text for a description of the circuit elements.
    \textbf{f,} Simulated resonance frequency $f_{\mathrm{r}}$ versus inverse of the MATBG kinetic inductance $1/L_{\mathrm{TBG}}$. Top axis represents $L_{\mathrm{TBG}}$.}
    \label{fig:fr-shift}
\end{figure}

The device is measured in a dilution refrigerator with a 20 mK base temperature ($T_{\mathrm{base}}$), and the effective temperature of the device with wiring connected is approximately 40 mK (see Methods and Supplementary Information for details). 
Figure~\ref{fig:fr-shift} shows the DC and microwave characterization of a representative MATBG device as a function of backgate voltage $V_\mathrm{BG}$. 
The differential resistance $\mathrm{d}V/\mathrm{d}I_{\mathrm{DC}}$ of the MATBG sample is first measured with a standard 4-probe lock-in technique at zero bias current (Figure~\ref{fig:fr-shift}). 
As expected, the device is resistive at the charge neutrality point (CNP) at filling factor $\nu=0$, and in the insulating regions around $\nu=\pm 2$ and $\nu=+3$. The DC resistance vanishes after passing the $\nu=\pm 2$ insulating regions, indicating the presence of superconductivity in both the hole-doped ($\nu<0$) and electron-doped ($\nu>0$) regimes (Fig.~\ref{fig:fr-shift}a).
We further parameterize the superconducting region by sweeping the DC bias current $I_{\mathrm{DC}}$ and backgate voltage $V_{\mathrm{BG}}$ in both the hole-doped (Fig.~\ref{fig:fr-shift}c) and the electron-doped (Fig.~\ref{fig:fr-shift}d) regimes.

We use a vector network analyzer (VNA) with ports 1 and 2 connected to the through-line to measure the microwave transmission coefficient $S_{21}$ as a function of frequency, from which the resonance frequency of the resonator is extracted~\autocite{Probst2015}. 
Figure~\ref{fig:fr-shift}b plots the magnitude $|S_{21}|$ as a function of frequency and back-gate voltage $V_{\mathrm{BG}}$. %
There are two notable features in the data: 
1) A constant resonant frequency $f_{\mathrm{r}}\approx 3.8~\mathrm{GHz}$ at the charge neutrality point 
$\nu=0$  and in the insulating regions at $\nu = \pm 2$, and 
2) a backgate-dependent resonance frequency in the superconducting regions, just beyond the insulating regions, $\nu \lesssim -2$ and $\nu \gtrsim 2$. 
%
The resonance frequency is generally gate-dependent within the superconducting regions defined via the DC-transport measurements (Fig.~\ref{fig:fr-shift}c-d). 

The presence and absence of resonator gate-dependence are modeled by the lumped-element circuits~\autocite{pozar2011microwave} in Fig.~\ref{fig:fr-shift}e, representing the coplanar waveguide (CPW) terminated by MATBG under the two following conditions. 
When the MATBG is highly resistive, \emph{i.e.,} at the CNP or in insulating regions, the MATBG termination is modeled as a resistor ($R_{\mathrm{TBG}}>>$ 1 k$\Omega$) in parallel with an inductive element $L_{\mathrm{prx}}$ (Fig.~\ref{fig:fr-shift}e, left schematic). 
The inductance $L_{\mathrm{prx}}$ represents the edge of the graphene that is in close proximity to the Ti/Al electrodes, from which the edge inherits a superconducting order parameter (see Methods: Microwave simulation for experimental confirmation of this model). 
The inductance $L_{\mathrm{prx}}$ of the proximitized graphene edge is essentially independent of gate voltage $V_{\mathrm{BG}}$-dependence due to the robust doping that results from the proximity effect and leading to a fixed resonance frequency (see Methods and Refs.~\cite{schmidt2018ballistic, wang2019coherent}). 
We establish $L_{\mathrm{prx}}=1.32$ nH by taking a sufficiently large resistance for the insulating state, $R_{\mathrm{TBG}}=100$ k$\Omega$, and performing microwave simulations to reproduce the observed nominal resonance frequency $f_{0}= 3.8$  GHz. 
In contrast, when the MATBG enters the superconducting phase, the resistance $R_{\mathrm{TBG}}$ vanishes and is replaced by a kinetic inductance $L_{\mathrm{TBG}}$ that arises from the superfluid condensate, resulting in a second parallel inductor that terminates the CPW and is $V_{\mathrm{BG}}$-dependent (Fig.~\ref{fig:fr-shift}e, right schematic). 
The resonance frequency $f_{\mathrm{r}}$ varies with gate voltage $V_{\mathrm{BG}}$ over the superconducting region determined by DC measurements (dashed lines, Fig.~\ref{fig:fr-shift}c, d), reaching maximum values around $V_{\mathrm{BG}}$=-6.8 V and 3 V within the hole-doped and electron-doped regimes, respectively (Fig.~\ref{fig:fr-shift}c, d).
Outside the insulating and superconducting regions, the MATBG bulk behaves like a normal metal with a small, non-zero resistance, which strongly damps the resonator and prevents an accurate extraction of its resonance frequency~\autocite{pozar2011microwave, Probst2015}.

Using this lumped-element model and incorporating the established value for $L_{\mathrm{prx}}$, we perform microwave simulations to extract the kinetic inductance of the MATBG termination given a measured resonance frequency of the resonator. 
Fig.~\ref{fig:fr-shift}f plots $f_{\mathrm{r}}$ as a function of $1/L_{\mathrm{TBG}}$.
The frequency shift $\Delta f_{\mathrm{r}}=f_{\mathrm{r}}-f_{0}$ depends almost linearly on $1/L_{\mathrm{TBG}}$ (see Methods for details).

We vary the refrigerator temperature to determine the MATBG critical temperature in the hole-doped and electron-doped regimes (Figs.~\ref{fig:stiff}a and \ref{fig:stiff}b). 
The critical temperature $T_{\mathrm{C}}^{(0.5)}$ -- the temperature where the resistance is half of the normal-state resistance -- is approximately 1 K across most of the gate-tunable area and consistent with previous reports on MATBG~\autocite{Cao2018} (see Supplementary Information for the definition of other critical temperatures $T_{\mathrm{C}}^{\mathrm{(onset)}}$ and $T_{\mathrm{C}}^{\mathrm{(zero)}}$). 

\subsection*{Superfluid Stiffness and the Quantum Metric in MATBG}
\begin{figure}[H]
    \centering
    \includegraphics[width=0.92\textwidth]{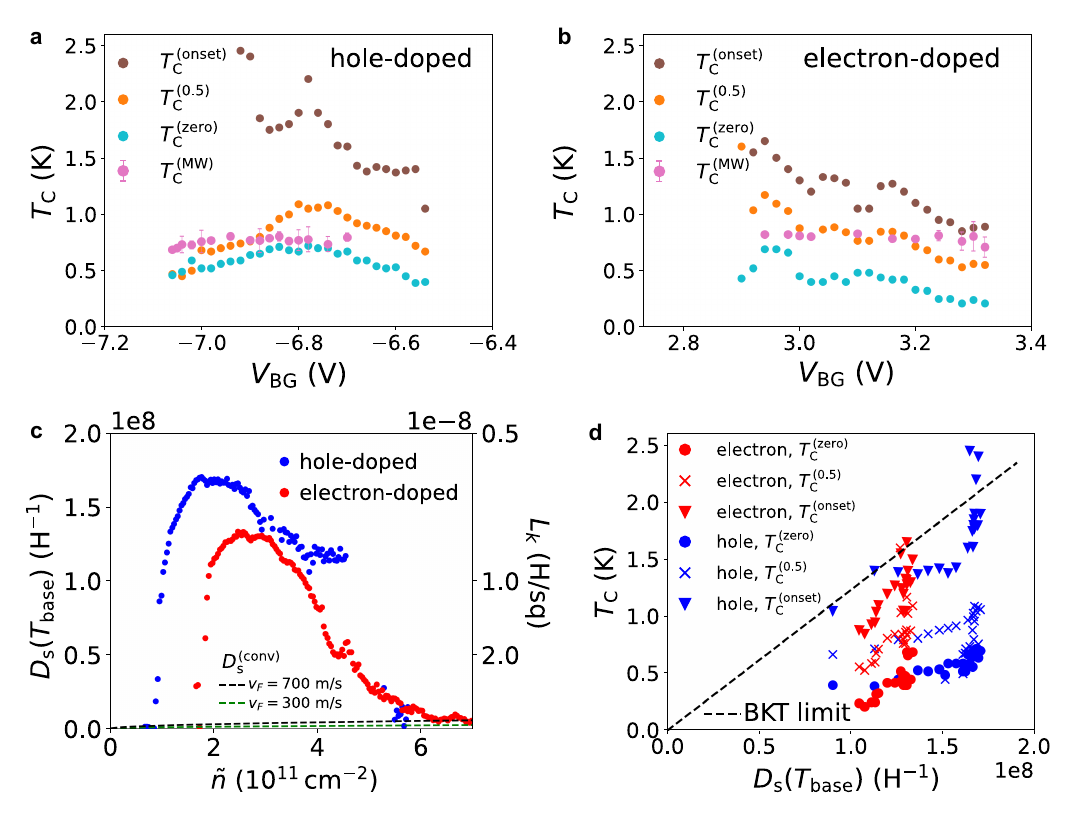}
    \caption{\textbf{Critical temperature and superfluid stiffness of superconducting MATBG.} 
    \textbf{a,b} Backgate dependence of critical temperatures 
    $T_{\mathrm{C}}^{\mathrm{(onset)}}$, $T_{\mathrm{C}}^{\mathrm{(0.5)}}$, and $T_{\mathrm{C}}^{\mathrm{(zero)}}$ as obtained from DC resistance measurements; and $T_{\mathrm{C}}^{\mathrm{(MW)}}$ as obtained from microwave measurements (see main text 
    for definitions of critical temperatures) in the hole-doped \textbf(a) and the electron-doped \textbf(b) regimes. 
    \textbf{c,} Superfluid stiffness $D_{\mathrm{s}}$ at base temperature $T_{\mathrm{base}}$ as a function of effective carrier density $\tilde{n}$, measured with respect to $|\nu|=2$. The error bars are too small to discern, with errors ranging from 1e6 to 5e6 $\mathrm{H}^{-1}$.
    The black and green dashed curves are the conventional contribution to the superfluid stiffness from Fermi liquid theory: $D_{\mathrm{s}}^{\mathrm{(conv)}}=e^{2}\tilde{n}v_{\mathrm{F}}/\hbar k_{\mathrm{F}}$ assuming $v_{\mathrm{F}}$=700 m/s and 300 m/s, respectively. 
    \textbf{d,} 
    Critical temperature $T_{\mathrm{C}}$ 
    and corresponding superfluid stiffness $D_{\mathrm{s}}$ at base temperature $T_{\mathrm{base}}$
    as tuned by $V_{\mathrm{BG}}$. 
    The black dashed line represents the BKT upper limit $T_{\mathrm{C}}=\pi\hbar^{2}D_{\mathrm{s}}(T_{\mathrm{base}})/8e^{2}k_{\mathrm{B}}$.}
    \label{fig:stiff}
\end{figure}

The kinetic inductance of MATBG enables us to directly measure the superfluid stiffness and its trends with physical parameters.
Assuming the active MATBG region has the same aspect ratio as the device itself, given by length $l$ and width $w$, the total measured superfluid stiffness $D_{\mathrm{s}}=l/(wL_{\mathrm{TBG}})$ at the base temperature $T_{\mathrm{base}}$ is plotted versus the effective carrier density in Fig.~\ref{fig:stiff}\textbf{c} for hole-doped (blue) and electron-doped (orange) MATBG. 

In 2D superconductors exhibiting a Berezinskii–Kosterlitz–Thouless (BKT) transition, $D_{\mathrm{s}}$ and $T_{\mathrm{C}}$ are related by the expression $\frac{\pi\hbar^{2}D_{\mathrm{s}}(0)}{8e^{2}}\geq k_{\mathrm{B}}T_{\mathrm{C}}$~\autocite{emery1995importance, kosterlitz1973ordering}. As illustrated in Fig.~\ref{fig:stiff}\textbf{d} (black dashed line),
the measured superfluid stiffness $D_{\mathrm{s}}(T_{\mathrm{base}})$ 
and the critical temperature $T_{\mathrm{C}}$ determined by DC resistance measurements exhibit a relationship that is generally consistent with this formula, providing additional confidence in our experimental measurement of $D_{\mathrm{s}}$.

We compare the measured $D_{\mathrm{s}}$ values to existing theoretical models. The conventional superfluid stiffness $D_{\mathrm{s}}^{\mathrm{(conv)}}$ predicted within the Fermi liquid framework yields 
$D_{\mathrm{s}}^{\mathrm{(conv)}}=e^{2}\tilde{n}v_{\mathrm{F}}/\hbar k_{\mathrm{F}}$, where $\tilde{n}$ is the effective carrier density measured relative to $\mid\nu\mid$=2, $k_{\mathrm{F}}=\sqrt{2\pi\tilde{n}}$ is the Fermi wave vector, and $v_{\mathrm{F}}$ is the Fermi velocity. From the $\mathrm{d}V/\mathrm{d}I$ versus $I_\mathrm{DC}$ measurements~\autocite{berdyugin2022out, tian2023evidence}, we estimate $v_{\mathrm{F}}$ to be in the range 300$\sim$700 m/s (see Supplementary Information). The dashed lines in Fig.~\ref{fig:stiff}\textbf{c} represent the calculated conventional stiffness $D_{\mathrm{s}}^{\mathrm{(conv)}}$ using these values.
The measured $D_{\mathrm{s}}$ is an order-of-magnitude greater than the conventional contribution $D_{\mathrm{s}}^{\mathrm{(conv)}}$. This discrepancy indicates mechanisms beyond Fermi-liquid theory need to be considered to fully describe the superconductivity in MATBG. 

The conventional superfluid stiffness $D_{\mathrm{s}}^{\mathrm{(conv)}}$ is proportional to the Fermi velocity, which becomes exceedingly small for a nearly flat energy band, resulting in a diminishing $D_{\mathrm{s}}^{\mathrm{(conv)}}$ in the flat-band limit and no superconductivity. 
It is proposed, however, that superfluidity and supercurrent can occur in a flat-band system if the band exhibits nontrivial quantum geometry~\autocite{peotta2015superfluidity}. This geometry manifests as overlapping Wannier functions between neighboring lattice sites, creating an extended state that enables the transport of interacting particles.

Since MATBG exhibits nearly flat bands at its Fermi level, the conventional contribution $D_{\mathrm{s}}^{\mathrm{(conv)}}$ is suppressed due to the corresponding small Fermi velocity. In this flat-band limit, the overall superfluid stiffness $D_{\mathrm{s}}$ is predominantly set by the quantum geometric contribution $D_{\mathrm{s}}^{\mathrm{(QG)}}$, which arises from delocalized Wannier functions influenced by the presence of higher energy bands~\autocite{hu2019geometric, julku2020superfluid}.

Our measurements indicate a maximum superfluid stiffness of $1.7 \times 10^{8}\ \mathrm{H}^{-1}$ for hole-doped MATBG and $1.25 \times 10^{8}\ \mathrm{H}^{-1}$ for electron-doped MATBG (Fig.~\ref{fig:stiff}c). 
Both are comparable with the theoretical prediction of $0.7\sim 5.2 \times 10^{8}\ \mathrm{H}^{-1}$ for $D_{\mathrm{s}}^{\mathrm{(QG)}}$~\autocite{hu2019geometric, julku2020superfluid} , suggesting that the dominant contribution to superfluid stiffness may arise from quantum geometry~\autocite{peotta2015superfluidity, xie2020topology, torma2022superconductivity, hu2019geometric, wu2019topological, julku2020superfluid,tian2023evidence, Tormarev2023, hu2019geometric}.

We also note that the superfluid stiffness directly measured in this experiment is consistent with the value extracted from measurements of the critical current density, as reported recently in Ref~\cite{tian2023evidence}.

\subsection*{Temperature Dependence of the MATBG Superfluid Stiffness}

The temperature dependence of the superfluid stiffness and the associated quasiparticle spectrum 
has been widely used to probe the gap anisotropy in unconventional superconductors~\autocite{prozorov2006magnetic, hardy2002magnetic}. 
At low temperatures ($T<0.3\,T_{\mathrm{C}}$), the conventional superfluid stiffness $D_{\mathrm{s}}^{\mathrm{(conv)}}$ in superconductors with isotropic gaps exhibits an exponential temperature dependence $\delta D_{\mathrm{s}}^{\mathrm{(conv)}}(T)/D_{\mathrm{s}}^{\mathrm{(conv)}}(0)\propto\sqrt{\frac{2\pi\Delta_{0}}{k_{\mathrm{B}}T}}\exp(-\frac{\Delta_{0}}{k_{\mathrm{B}}T})$, where $\delta D_{\mathrm{s}}^{\mathrm{(conv)}}(T)=D_{\mathrm{s}}^{\mathrm{(conv)}}(0)-D_{\mathrm{s}}^{\mathrm{(conv)}}(T)$ is the change in stiffness with temperature $T$, and $\Delta_{0}$ is the superconducting gap at zero temperature~\autocite{prozorov2006magnetic, tinkham2004introduction,}. 
In contrast, superconductors with an anisotropic gap or nodes exhibit a power-law temperature dependence $\delta D_{\mathrm{s}}^{\mathrm{(conv)}}(T)/D_{\mathrm{s}}^{\mathrm{(conv)}}(0)\propto T^{n}$ ~\autocite{prozorov2006magnetic}. 
In 2D momentum space, nodal gaps exhibit an exponent of $n=1$ (linear dependence) in the clean limit and $n=2$ in the dirty limit,
while nodeless anisotropic gaps display $n>2$~\autocite{hirschfeld1993effect}. 
The isotropic s-wave gap with an exponential dependence typically displays $n>4$ in this formalism~\autocite{roppongi2023bulk, teknowijoyo2018nodeless}.
On the other hand, recent theoretical studies suggest that the quantum geometric contribution $D_{\mathrm{s}}^{\mathrm{(QG)}}$ follows a power-law temperature dependence, with the value of the exponent potentially depending on the structure of the superconducting gap~\autocite{penttilä2024flatband}.

\begin{figure}[H]
    \centering
    \includegraphics[width=0.93\textwidth]{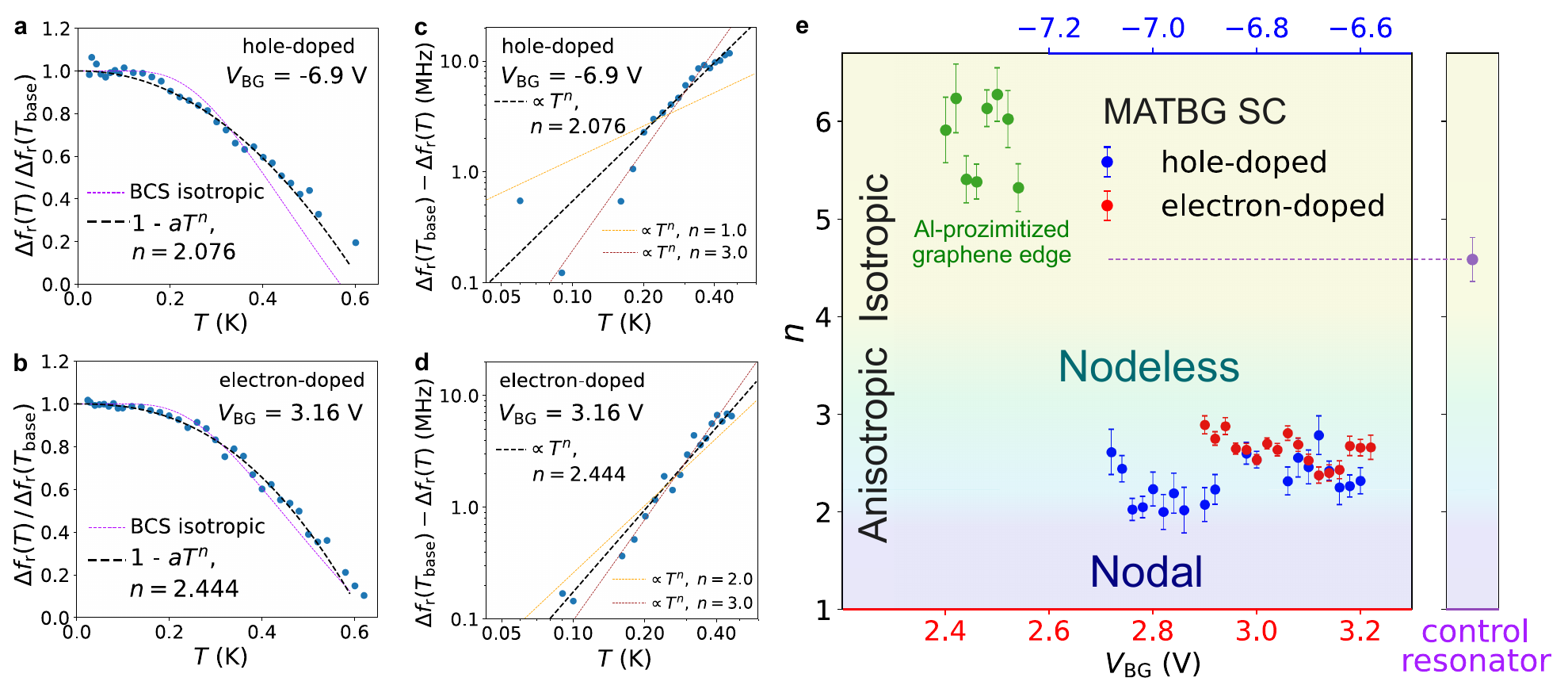}
    \caption{\textbf{Temperature-dependent shift in resonant frequency due to varying superfluid stiffness.} 
    \textbf{a,b, } Temperature dependence of $\Delta f_{\mathrm{r}}(T)/\Delta f_{\mathrm{r}}(T_{\mathrm{base}})$ for $T<T_{\mathrm{BKT}}$ (See Supplementary Information for determination of $T_{\mathrm{BKT}}$.) in the hole-doped (\textbf{a}) and the electron-doped (\textbf{b}) regimes. Blue dots represent the experimental data, black dashed lines depict the power-law fitting, and purple dashed lines depict the exponential function in BCS isotropic model. 
    \textbf{c,d,} Log-log plot of $\Delta f_{\mathrm{r}}(T_{\mathrm{base}})-\Delta f_{\mathrm{r}}(T)$ in the hole-doped (\textbf{c}) and the electron-doped (\textbf{d}) regimes. Orange and brown dashed lines depict the power-law with exponent of n=1.0, 4.0 and 2.0, 4.0, respectively. 
    \textbf{e,} Power-law fit exponent of temperature dependence of the resonance in the hole-doped superconducting region (blue dots), electron-doped superconducting region (red dots), insulating region where the Al-proximitized graphene terminate the resonator (green dots), and the control resonator (purple dot in the right panel).}
    \label{fig:Tdependence}
\end{figure}

To test these ideas, we study the temperature dependence of the MATBG kinetic inductance. 
Figure~\ref{fig:Tdependence} shows the temperature dependence of the resonant frequency in the superconducting phase for both the hole-doped and electron-doped regimes. 
The change in frequency $\Delta f_{\mathrm{r}}=f_{\mathrm{r}}(V_{\mathrm{BG}})-f_{0}$ is proportional to the 
superfluid stiffness (see Fig.~\ref{fig:fr-shift}f) and decreases with temperature. 
It eventually reaches zero around the critical temperature determined from DC resistance measurements. 
Note that the frequency shift attributed to the geometric inductance of the Al resonator and the proximitized graphene $L_{\mathrm{prx}}$ is considerably smaller than the $V_{\mathrm{BG}}$-dependent frequency shift due to $L_{\mathrm{TBG}}$ (see Supplementary Information). 

Figures~\ref{fig:Tdependence}a and \ref{fig:Tdependence}b show $\Delta f_{\mathrm{r}}(T)\,/\,\Delta f_{\mathrm{r}}(T_{\mathrm{base}})$ -- equivalent to $D_{\mathrm{s}}(T)/D_{\mathrm{s}}(T_{\mathrm{base}})$ -- for temperatures $T<T_{\mathrm{BKT}}$, where $T_{\mathrm{BKT}}$ is the Berezinskii–Kosterlitz–Thouless temperature (see Supplementary Information for the determination of $T_{\mathrm{BKT}}$) in the hole-doped ($V_{\mathrm{BG}}=-6.9$ V) and electron-doped ($V_{\mathrm{BG}}=3.16$ V) regimes. 
The trends clearly deviate from the exponential dependence expected for a BCS isotropic model, indicating unconventional superconductivity in MATBG.
Furthermore, they fit well to a power-law function with exponent \textit{n}= 2.08 and \textit{n}= 2.44, respectively. 
The power-law dependence is also confirmed in the logarithmic plot of $\Delta f_{\mathrm{r}}(T_{\mathrm{base}})-\Delta f_{\mathrm{r}}(T)\propto \delta D_{\mathrm{s}}(T)$
depicted in Figures~\ref{fig:Tdependence}c and \ref{fig:Tdependence}d. 
We extract the power-law exponent across the entire superconducting dome in both electron- and hole-doped regimes, with $n$ ranging from 2 to 3 (Fig.~\ref{fig:Tdependence}e). This suggest an anisotropic superconducting gap within the Fermi liquid framework $D_{\mathrm{s}}^{\mathrm{(conv)}}$~\autocite{prozorov2006magnetic, hardy2002magnetic}. In contrast, when the same measurement and analysis is performed in the insulating MATBG regime, where the $\lambda/4$ resonator is terminated only by the Al-proximitized graphene edge, or on the Al control resonator, we observe higher exponents ($n \geq 4.6$), characteristic of isotropic s-wave superconductivity. 

Notably, this power-law behavior with $n$ ranging from 2 to 3 also align with theoretical predictions for $D_{\mathrm{s}}^{\mathrm{(QG)}}$ in  a superconductor with an anisotropic gap\autocite{penttilä2024flatband}. While our results strongly indicate an anisotropic gap independent of the model used, further investigation is required to fully interpret these results to better understand the gap structure and pairing symmetry in MATBG. Finally, we note that other extrinsic 
disorder may also lead to anisotropic or even nodal gap \autocite{khvalyuk2024near, ghosal2001inhomogeneous}, or, conversely, lift the unprotected gap nodes\autocite{mishra2009lifting, cho2016energy}
. The data from the additional sample showing similar results are discussed in the Supplementary Information.

\subsection*{Bias Current and Microwave Power Dependence}
\begin{figure}[H]
    \centering
    \includegraphics[width=0.96\textwidth]{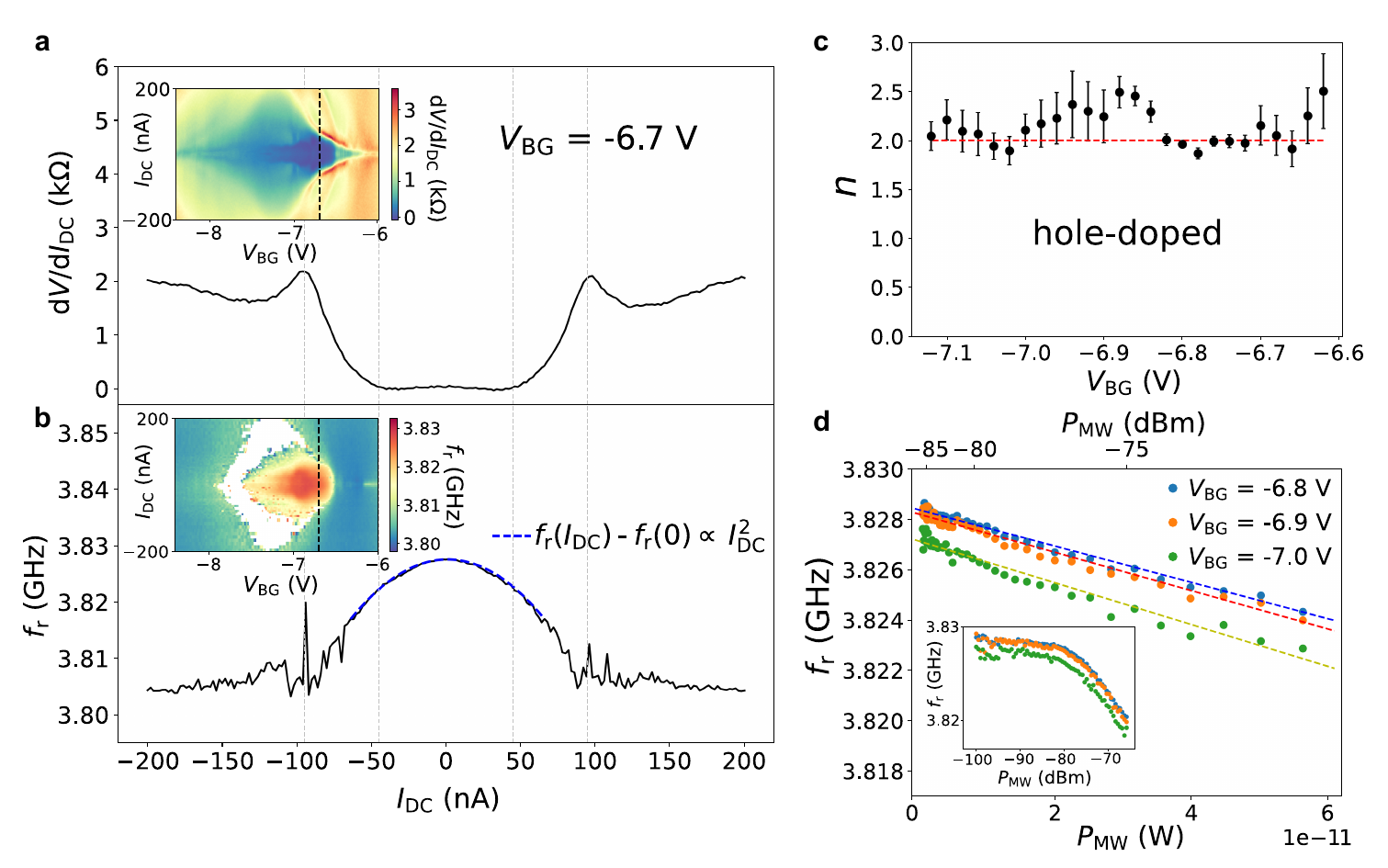}
    \caption{\textbf{DC bias and microwave power dependence of resonant frequency.} 
    \textbf{a,} Differential resistance $\mathrm{d}V/\mathrm{d}I_{\mathrm{DC}}$ dependence on bias current $I_{\mathrm{DC}}$ for backgate voltage  $V_{\mathrm{BG}}=-6.7$ V (black dashed line in the inset). Inset: 2D color map versus $V_{\mathrm{BG}}$ and $I_{\mathrm{DC}}$. 
    \textbf{b,} Resonant frequency $f_{\mathrm{r}}$ dependence on $I_{\mathrm{DC}}$ for backgate voltage  $V_{\mathrm{BG}}=-6.7$ V (black dashed line in the inset). Inset: 2D color map of dV/dI versus $V_{\mathrm{BG}}$ and $I_{\mathrm{DC}}$. The blue dashed curve indicates a quadratic fit to the data. 
    \textbf{c,} Exponents $n$ of the power-law fitting $f_{\mathrm{r}}(I_{\mathrm{DC}}=0)-f_{\mathrm{r}}(I_{\mathrm{DC}})\propto I_{\mathrm{DC}}^{n}$ as a function of $V_{\mathrm{BG}}$. Red dashed line indicates $n=2$.
    \textbf{d,} Microwave power $P_{\mathrm{MW}}$ dependence of $f_{\mathrm{r}}$ at $V_{\mathrm{BG}}$=-6.8, -6.9, and -7.0 V. Blue dashed lines indicate a linear dependence with power, corresponding to a quadratic dependence on microwave current amplitude.}
    \label{fig:Biasdependence}
\end{figure}

We now explore the dependence of the superfluid stiffness on DC bias current. 
We apply a DC bias current $I_{\mathrm{DC}}$ through terminal $I_{+}$ (Fig.~\ref{fig:sample}c). 
Figures~\ref{fig:Biasdependence}a and ~\ref{fig:Biasdependence}b respectively display the DC differential resistance $\mathrm{d}V/\mathrm{d}I_{\mathrm{DC}}$ and the resonant frequency $f_{\mathrm{r}}$ versus current $I_{\mathrm{DC}}$ at $V_{\mathrm{BG}}=-6.7\,$V. 
Importantly, $f_{\mathrm{r}}$ shows a noticeable dependence over the range of currents $0\leq I_{\mathrm{DC}}\leq 40$ nA, even around zero bias, whereas the DC resistance consistently remains a constant 0 $\Omega$. 
The fact that only the resonance frequency changes with $I_{\mathrm{DC}}$ while the sample remains superconducting indicates that it is the superfluid stiffness that changes ($D_{\mathrm{s}}$ is suppressed, $L_{\mathrm{K}}$ increases) with increasing $I_{\mathrm{DC}}$. At any given gate voltage within the superconducting domes, the frequency shift exhibits a quadratic dependence (concave down) on the DC bias current (blue dashed line in Fig.~\ref{fig:Biasdependence}\textbf{b} and \textbf{c}).

In addition, the resonant frequency $f_{\mathrm{r}}$ decreases linearly with the microwave power $P_{\mathrm{MW}}$ applied to the feedline (Fig. \ref{fig:Biasdependence}d). 
Considering that the $P_{\mathrm{MW}}$ is proportional to the square of the AC current amplitude, $P_{\mathrm{MW}}\propto I_{\mathrm{MW}}^{2}$, this similarly indicates that the reduction of $f_{\mathrm{r}}$ depends quadratically on microwave current. 
Note that the power dependence of the MATBG-terminated resonator is driven primarily by the intrinsic MATBG response, as the contribution to the frequency shift from the Al resonator and the proximitized edges of the MATBG is negligibly small (see Methods and Extended figures 3 and 4).

If spatial inversion symmetry is preserved in this system, second-order nonlinearity is forbidden, so the leading correction to the inductive response occurs at third order, resulting in a second-order modification of the inductance and superfluid stiffness with respect to the current amplitude. This behavior is described by Ginzburg-Landau theory and by BCS theory when considering pair-breaking by the finite momentum of the Cooper pairs. It is expressed as $D_{\mathrm{s}}(I)/D_{\mathrm{s}}(0)=1-(I/I^{*})^{2}$ or $L_{\mathrm{K}}(I)\propto L_{0} [1+(I/I^{*})^{2}]$, where $I^{*}$ is on the order of the depairing current~\autocite{gittleman1965nonlinear, enpuku1993modulation,kubo2020superfluid,anthore2003density,}, and is commonly observed in thin film superconductors~\autocite{ho2012wideband, luomahaara2014kinetic, Ku2010, Ciaassen1999, Thomas2020, Vissers2015, Kirsh2021, Zhao2020}. To our knowledge, the DC bias current dependence of superfluid stiffness arising from the quantum geometry contribution, $D_{\mathrm{s}}^{\mathrm{(QG)}}$, has not yet been discussed theoretically.

If the superconducting gap is nodal, the quadratic dependence may be replaced by a linear dependence above a current threshold set by the disorder~\autocite{Dahm1999, yip1992nonlinear, dahm1996theory, Lee2005, Wilcox2022,}, known as the nonlinear Meissner effect. However, it is important to note that the nonlinear Meissner effect is not necessarily a universally observed characteristic of nodal superconductors~\autocite{carrington1999absence, li1998nonlinear}(see Supplementary Information). For example, this effect was recently observed in a spatially inhomogeneous superconducting film with a nodeless gap, reflecting quasiparticle excitation arising from weak links~\autocite{Makita2022}.
Here, our devices do not exhibit this nonlinear Meissner effect; the quadratic dependence on current holds up to $I_{\mathrm{C}}$ throughout the superconducting dome.

\subsection*{Discussion and Conclusion}
We develop and demonstrate a measurement platform comprising a microwave resonator terminated by a sample of interest -- in this work, MATBG -- that is contacted to enable both DC and microwave measurements. The resonator with MATBG termination features a quality factor exceeding 1000 at 20 mK (see Supplementary Imformation), capable of discerning changes in the superfluid stiffness at the 1\% level. 
We use this platform to measure the MATBG kinetic inductance -- a direct measurement of the superfluid stiffness -- as a function of parameters that lead to pair-breaking, including thermal excitation, bias current, and microwave drive power. 

The directly measured MATBG superfluid stiffness is an order-of-magnitude larger than the conventional contribution 
predicted by the band dispersion within a Fermi liquid framework. 
Instead, this value aligns well with a theoretical model that incorporates quantum geometry, where a $D_{\mathrm{s}}$ of this magnitude for MATBG is expected~\autocite{peotta2015superfluidity, xie2020topology, torma2022superconductivity, hu2019geometric, wu2019topological, julku2020superfluid, Tormarev2023, penttilä2024flatband}. This result, along with recent DC transport measurements~\autocite{tian2023evidence}, underscores the potentially significant role of quantum geometry in understanding MATBG superconductivity.

The temperature dependence of the MATBG superfluid stiffness follows a power-law dependence in both the hole-doped and electron-doped regimes. This behavior deviates from the exponential dependence observed in superconductors with isotropic s-wave pairing as well as the linear dependence observed in clean nodal superconductors.
Our data generally falls within the range of power-law exponents $n=2\ldots 3$.

Within the conventional Fermi liquid framework, the measured exponents indicate the presence of an anisotropic gap in MATBG~\autocite{prozorov2006magnetic, hardy2002magnetic}. 
The exponents, together with the quadratic bias-current dependence of $D_{\mathrm{s}}$ -- which indicates the absence of a nonlinear Meissner effect in our measurements -- would suggest a nodeless anisotropic gap in MATBG.
We do not rule out the possibility that our results may reflect an intrinsically nodal gap that is lifted due to the presence of disorder, e.g., from twist-angle variations.

However, the 
contribution of quantum geometry to the low-energy quasiparticle spectrum remains an open question, so it is not guaranteed that the conventional analysis based on power-law exponents would apply precisely. 
Within a quantum-geometry framework, recent theoretical calculations for anisotropic gaps that include a quantum geometric contribution predict an exponent of 3~\autocite{penttilä2024flatband}, which is consistent with our measured exponents.

In summary, we develop and demonstrate a platform that directly measures the superfluid stiffness of superconducting systems. Our experimental measurements of the superfluid stiffness in MATBG reveal phenomena that apparently go beyond the conventional Fermi-liquid theory and strongly suggest a connection to multi-band quantum geometry in a flat-band superconductor. 
While it is beyond the scope of this experiment to distinguish the relative contributions of the Fermi-liquid and quantum geometry frameworks to the superfluid stiffness, our results indicate anisotropic pairing in MATBG within both frameworks. 
These findings lay the groundwork for future theoretical studies to deepen our understanding of the underlying physics.

During the preparation of our manuscript, we became aware of a complementary work by Banerjee and Hao \textit{et al.}, from Philip Kim's team at Harvard~\autocite{banerjee2024}.

\section*{Methods}
\subsection*{Device fabrication}
We begin with a 250 nm aluminum film deposited on high-resistivity Si wafers. The circuit elements, including a transmission line, resonators, LC filters, and DC probe lines (Fig. 1a and 1b), are patterned using photolithography and an aluminum wet-etch process. A Ti/Al (5/30 nm) back gate is defined using E-beam lithography and e-beam evaporation within the cutout window used to host the vdW heterostructures. Details of the aluminum fabrication process can be found in ref.~\cite{yan2016flux}.

The incorporation of vdW heterostructures onto the patterned chips employs standard mechanical exfoliation and dry polymer-based techniques~\autocite{wang2019coherent, wang2013one}. The \ch{hBN} flake exfoliated on the \ch{SiO2} substrate is transferred onto a Ti/Al (5/30 nm) back gate. After removing the polymer by soaking in chloroform, the top surface of the \ch{hBN} is cleaned by contact-mode AFM scanning. Following an iterative stacking procedure, a stack of \ch{hBN} and magic angle twisted bilayer graphene (MATBG) is released onto the bottom \ch{hBN}.  

After transferring the heterostructures, the hBN/MATBG/hBN stack is patterned into a Hall bar geometry using E-beam lithography and reactive ion etching (RIE). Superconducting contacts are made to the MATBG edge using RIE followed by thermal evaporation of Ti/Al. The superconducting bridging between the contacts and the aluminum resonator and the ground plane is made using E-beam lithography and in-situ ion-milling followed by aluminum deposition. 

\subsection*{Measurement setup}
The experiment is performed in a Bluefors XLD-1000 dilution refrigerator with a base temperature of approximately 20 mK. Attenuation at several cryogenic stages 
is used to reduce the number of thermal photons at higher-temperature stages from reaching the device. 
We use a -20 dB directional coupler to pump a Josephson travelling wave parametric amplifier (TWPA) mounted at base temperature to pre-amplify the resonator probe tone. To reduce the impact of reflections of the TWPA pump on the device, we add a circulator between the device and the TWPA. Following the TWPA, there are additional isolators, filters, and a high-electron mobility transistor (HEMT) amplifier (LNF) thermally anchored to the 3K stage. At room temperature (300K), we further amplify (MITEQ) the output signal. Each DC line connected to the device is lowpass filtered with a $\pi$-filter at 3 K, and a QFilter (Quantum Machines) at the mixing chamber stage. See \ref{fig:wiring} for a schematics of the measurement setup.

\subsection*{Microwave simulation}
We use the SONNET finite-element solver to determine the relationship between the resonant frequency shift and the kinetic inductance. We model the Al film as a two-dimensional perfect electrical conductor on a 350 $\mathrm{\mu m}$ Silicon (Si) substrate , with 500 um vacuum above the film, and calculate the S-parameters at the microwave ports.

Our device model comprises an inductor -- representing the always-superconducting proximitized edge of the graphene near the Ti/Al contacts -- in parallel with the gate-voltage-dependent resistive or inductive impedance of the MATBG (see main text). The proximitized edge accounts for the fact that changing the MATBG from superconducting to insulating does not change the nature of the resonator from lambda/4 to lambda/2, as might be naively expected. We test the validity of this model with two additional experiments.

First, we fabricate a Bernal (AB)-stacked bilayer graphene device that nominally has the same geomtery and metalization as the MATBG device. The Bernal device exhibits a resonance around 4 GHz when biased near the CNP to be highly insulating (\ref{fig:simulation}b). Since the Bernal-stacked bilayer graphene is non-superconducting at this bias point, this observation indicates that the observed resonance around 4 GHz is not unique to the MATBG sample and also unrelated to its intrinsic superconductivity.

Second, we verify this model by applying a perpendicular magnetic field to the Bernal (AB)-stacked bilayer graphene device. \ref{fig:simulation}b shows the magnetic field dependence of the resonance in the bilayer graphene device gate-biased at the CNP. The resonance at 4.24 GHz vanishes at around 0.42 mT, which is much smaller than the critical magnetic field of either the resonator aluminum film ($H_{\mathrm{C}}\approx$ 10 mT) or the superconducting MATBG, but consistent with a weak superconducting link to ground formed by proximitized graphene~\autocite{bretheau2017tunnelling}. These tests verify the role of the proximitized graphene in maintaining a $\lambda/4$ resonator with a resonance frequency that only shifts about 1\% between insulating and superconducting MATBG regimes.

To analyze the lumped-element model, we set the resistance $R_{\mathrm{TBG}}$=100 k$\Omega$ to represent the insulating regime and sweep $L_{\mathrm{prx}}$ to reproduce the observed base frequency of 3.8 GHz (\ref{fig:simulation}d inset). This yields $L_{\mathrm{prx}}$=1.32 nH, consistent with a previous study of the Josephson inductance in proximitized graphene~\autocite{schmidt2018ballistic,}. Using this value, we calculate the $L_{\mathrm{TBG}}$ dependence of the $f_{\mathrm{r}}$ (\ref{fig:simulation}d). The frequency shift $\Delta f_{\mathrm{r}}=f_{\mathrm{r}}-3.8$ GHz depends almost linearly on $1/L_{\mathrm{TBG}}\propto D_{\mathrm{s}}$.

Throughout this simulation, the capacitance between the MATBG and the back gate is fixed at 3 fF, a value estimated based on the dimension and dielectric properties of the bottom hBN. We find that the resonance frequency is insensitive to this capacitance (\ref{fig:simulation}c). Also, we check the $L_{\mathrm{TBG}}$ dependence of $f_{\mathrm{r}}$ with several different values of $L_{\mathrm{prx}}$, as shown in Extended Figures 2d and e. The linear dependence between $\Delta f_{\mathrm{r}}$ and $1 / L_{\mathrm{TBG}}$ holds for different values of $L_{\mathrm{prx}}$, which justifies the analysis assuming $\Delta f_{\mathrm{r}}\propto D_{\mathrm{s}}$.

To illustrate the impact of possible normal regions caused by contact resistance, disorder, and twist-angle inhomogeneity in the MATBG device, we perform simulations incorporating a finite resistance in parallel and series configurations. As shown in Extended Figures 2f and g, the presence of normal regions affects the quality factor but does not result in any discernable frequency shift, thereby not impacting the measurement of $L_{\mathrm{TBG}}$ (\ref{fig:simulation}f and g).      

\subsection*{Aluminum control resonator and resonance in the insulating phase of MATBG}
We characterize the aluminum control-resonator in the absence of MATBG to test our protocol for measuring the superfluid stiffness and to confirm that the MATBG-terminated resonator temperature and power dependence is dominated by the MATBG.

\ref{fig:cont}b plots the resonance frequency of the $\lambda/4$ aluminum control resonator terminated directly to ground (i.e., without MATBG). Since the resonant frequency is  $f_{\mathrm{r}}=\frac{1}{2\pi\sqrt{(L_{\mathrm{G}}+L_{\mathrm{K}})C}}$, where $L_{\mathrm{G}}\gg L_{\mathrm{K}}$ is geometric inductance, the relation $f_{\mathrm{r}}(T_{\mathrm{base}})\,/\,f_{\mathrm{r}}(T)-1$ is proportional to  $L_{\mathrm{K}}$. 
The behavior is well fitted by the conventional isotropic BCS Fermi liquid model, consistent with the conventional superconductivity present in aluminum (see \ref{fig:cont}c). The power-law fit gives an exponent of 4.585, consistent with previous reports for aluminum and isotropic gaps in general~\autocite{roppongi2023bulk, teknowijoyo2018nodeless}. 
The shift in resonance frequency due to the kinetic inductance in the aluminum-only control resonator is much smaller than the MATBG-terminated aluminum experiment resonator, in part because the of Al (250 nm) is much larger than that of MATBG ($\sim$1nm). 
This means the geometric inductance dominates in the aluminum-only resonator. 
The microwave power dependence of $f_{\mathrm{r}}$ is negligible compared with that in the experimental MATBG resonator (\ref{fig:cont}d).

Extended Figures 4b and 4c presents the temperature dependence of the resonance for the MATBG-terminated resonator measured at the backgate voltage of $V_{\mathrm{BG}}=2.44$ V in insulating region near $\nu=2$. Here, the termination is shunted by the the Al-prozimitized graphene edge.
The temperature dependence remains flat below 0.5 K, in contrast to the resonance when the MATBG is biased in the superconducting phase, and is fitted by the isotropic BCS model using the $T_{\mathrm{C}}$=1.2 K. Here also, the power-law exponent n=5.405 is consistent with an isotropic gap in the proximitized region \autocite{roppongi2023bulk, teknowijoyo2018nodeless}.
We used this resonance frequency as a standard reference when analyzing the frequency shift due to the MATBG kinetic inductance: 
$\Delta f_{\mathrm{r}}(V_{\mathrm{BG}}, T)=f_{\mathrm{r}}(V_{\mathrm{BG}}, T)-f_{\mathrm{r}}(V_{\mathrm{BG}}=\mathrm{2.44V}, T)$.

The microwave power dependence of $f_{\mathrm{r}}$ is negligible in the insulating region compared with that in the superconducting region (\ref{fig:insu}d). 
This indicates that the power dependence presented in the main text for the MATBG-terminated resonator is due primarily to the MATBG superconductivty. 

\printbibliography

@article{banerjee2024,
  title={Superfluid stiffness of twisted multilayer graphene superconductors}, 
      author={Abhishek Banerjee and Zeyu Hao and Mary Kreidel and Patrick Ledwith and Isabelle Phinney and Jeong Min Park and Andrew M. Zimmerman and Kenji Watanabe and Takashi Taniguchi and Robert M Westervelt and Pablo Jarillo-Herrero and Pavel A. Volkov and Ashvin Vishwanath and Kin Chung Fong and Philip Kim},
  journal={arXiv preprint arXiv:2406.13742},
  year={2024}
}

@article{andrei2021marvels,
  title={The marvels of moir{\'e} materials},
  author={Andrei, Eva Y and Efetov, Dmitri K and Jarillo-Herrero, Pablo and MacDonald, Allan H and Mak, Kin Fai and Senthil, T and Tutuc, Emanuel and Yazdani, Ali and Young, Andrea F},
  journal={Nature Reviews Materials},
  volume={6},
  number={3},
  pages={201--206},
  year={2021},
  publisher={Nature Publishing Group UK London}
}

@article{kannan2023demand,
  title={On-demand directional microwave photon emission using waveguide quantum electrodynamics},
  author={Kannan, Bharath and Almanakly, Aziza and Sung, Youngkyu and Di Paolo, Agustin and Rower, David A and Braum{\"u}ller, Jochen and Melville, Alexander and Niedzielski, Bethany M and Karamlou, Amir and Serniak, Kyle and others},
  journal={ Nat. Phys.},
  volume={19},
  number={3},
  pages={394--400},
  year={2023},
  publisher={Nature Publishing Group UK London}
}

@article{Koch2007,
  title = {Charge-insensitive qubit design derived from the Cooper pair box},
  author = {Koch, Jens and Yu, Terri M. and Gambetta, Jay and Houck, A. A. and Schuster, D. I. and Majer, J. and Blais, Alexandre and Devoret, M. H. and Girvin, S. M. and Schoelkopf, R. J.},
  journal = {Phys. Rev. A},
  volume = {76},
  issue = {4},
  pages = {042319},
  numpages = {19},
  year = {2007},
  
  publisher = {American Physical Society},
  doi = {10.1103/PhysRevA.76.042319},
 % url = {https://link.aps.org/doi/10.1103/PhysRevA.76.042319}
}

@article{Hoi2011,
  title = {Demonstration of a Single-Photon Router in the Microwave Regime},
  author = {Hoi, Io-Chun and Wilson, C. M. and Johansson, G\"oran and Palomaki, Tauno and Peropadre, Borja and Delsing, Per},
  journal = {Phys. Rev. Lett.},
  volume = {107},
  issue = {7},
  pages = {073601},
  numpages = {5},
  year = {2011},
  
  publisher = {American Physical Society},
  doi = {10.1103/PhysRevLett.107.073601},
 % url = {https://link.aps.org/doi/10.1103/PhysRevLett.107.073601}
}

@article {Astafiev2010,
	author = {Astafiev, O. and Zagoskin, A. M. and Abdumalikov, A. A. and Pashkin, Yu. A. and Yamamoto, T. and Inomata, K. and Nakamura, Y. and Tsai, J. S.},
	title = {Resonance Fluorescence of a Single Artificial Atom},
	volume = {327},
	number = {5967},
	pages = {840--843},
	year = {2010},
	doi = {10.1126/science.1181918},
	publisher = {American Association for the Advancement of Science},
	abstract = {The coherence properties of superconducting circuits enable them to be developed as qubits in quantum information processing applications. Astafiev et al. (p. 840) now show that these macroscopic superconducting devices also behave as artificial atoms and can exhibit quantum optical effects. The ability to fabricate and integrate these superconducting devices in electronic circuitry may help toward developing a fully controlled quantum optics system on a chip.An atom in open space can be detected by means of resonant absorption and reemission of electromagnetic waves, known as resonance fluorescence, which is a fundamental phenomenon of quantum optics. We report on the observation of scattering of propagating waves by a single artificial atom. The behavior of the artificial atom, a superconducting macroscopic two-level system, is in a quantitative agreement with the predictions of quantum optics for a pointlike scatterer interacting with the electromagnetic field in one-dimensional open space. The strong atom-field interaction as revealed in a high degree of extinction of propagating waves will allow applications of controllable artificial atoms in quantum optics and photonics.},
	issn = {0036-8075},
	%URL = {https://science.sciencemag.org/content/327/5967/840},
	journal = {Science}
}

@article{Hoi2013,
	doi = {10.1088/1367-2630/15/2/025011},
	%url = {https://doi.org/10.1088%2F1367-2630%2F15%2F2%2F025011},
	year = 2013,
	
	publisher = {{IOP} Publishing},
	volume = {15},
	number = {2},
	pages = {025011},
	author = {Io-Chun Hoi and C M Wilson and Göran Johansson and Joel Lindkvist and Borja Peropadre and Tauno Palomaki and Per Delsing},
	title = {Microwave quantum optics with an artificial atom in one-dimensional open space},
	journal = {New Journal of Physics},
	abstract = {We address recent advances in microwave quantum optics with artificial atoms in one-dimensional (1D) open space. This field relies on the fact that the coupling between a superconducting artificial atom and propagating microwave photons in a 1D open transmission line can be made strong enough to observe quantum coherent effects, without using any cavity to confine the microwave photons. We investigate the scattering properties in such a system with resonant coherent microwaves. We observe the strong nonlinearity of the artificial atom and under strong driving we observe the Mollow triplet. By applying two resonant tones, we also observe the Autler–Townes splitting. Exploiting these effects, we demonstrate two quantum devices at the single-photon level in the microwave regime: the single-photon router and the photon-number filter. These devices provide important steps toward the realization of an on-chip quantum network.}
}

@Article{Mirhosseini2019,
author={Mirhosseini, Mohammad
and Kim, Eunjong
and Zhang, Xueyue
and Sipahigil, Alp
and Dieterle, Paul B.
and Keller, Andrew J.
and Asenjo-Garcia, Ana
and Chang, Darrick E.
and Painter, Oskar},
title={Cavity quantum electrodynamics with atom-like mirrors},
journal={Nature},
year={2019},
volume={569},
number={7758},
pages={692-697},
doi={10.1038/s41586-019-1196-1},
%url={https://doi.org/10.1038/s41586-019-1196-1}
}

@article{balents2020superconductivity,
  title={Superconductivity and strong correlations in moir{\'e} flat bands},
  author={Balents, Leon and Dean, Cory R and Efetov, Dmitri K and Young, Andrea F},
  journal={Nature Physics},
  volume={16},
  number={7},
  pages={725--733},
  year={2020},
  publisher={Nature Publishing Group UK London}
}

@book{pozar2011microwave,
  title={Microwave engineering},
  author={Pozar, David M},
  year={2011},
  publisher={John wiley \& sons}
}

@article{hirschfeld1993effect,
  title={Effect of strong scattering on the low-temperature penetration depth of a d-wave superconductor},
  author={Hirschfeld, Peter J and Goldenfeld, Nigel},
  journal={Phys. Rev. B},
  volume={48},
  number={6},
  pages={4219},
  year={1993},
  publisher={APS}
}

@article{roppongi2023bulk,
  title={Bulk evidence of anisotropic s-wave pairing with no sign change in the kagome superconductor CsV3Sb5},
  author={Roppongi, M and Ishihara, K and Tanaka, Y and Ogawa, K and Okada, K and Liu, S and Mukasa, K and Mizukami, Y and Uwatoko, Y and Grasset, R and others},
  journal={Nat. Commun.},
  volume={14},
  number={1},
  pages={667},
  year={2023},
  publisher={Nature Publishing Group UK London}
}

@article{Ku2010,
  title={Superconducting nanowires as nonlinear inductive elements for qubits},
  author={Ku, Jaseung and Manucharyan, Vladimir and Bezryadin, Alexey},
  journal={Phys. Rev. B},
  volume={82},
  number={13},
  pages={134518},
  year={2010},
  publisher={APS}
}

@article{Ciaassen1999,
  title={Large non-linear kinetic inductance in superconductor/normal metal bilayer films},
  author={Claassen, JH and Adrian, S and Soulen, RJ},
  journal={IEEE Trans. Appl. Supercond.},
  volume={9},
  number={2},
  pages={4189--4192},
  year={1999},
  publisher={IEEE}
}

@article{Thomas2020,
  title={Nonlinear effects in superconducting thin film microwave resonators},
  author={Thomas, Christopher N and Withington, Stafford and Sun, Zhenyuan and Skyrme, Tess and Goldie, David J},
  journal={New J. Phys.},
  volume={22},
  number={7},
  pages={073028},
  year={2020},
  publisher={IOP Publishing}
}

@article{Vissers2015,
  title={Frequency-tunable superconducting resonators via nonlinear kinetic inductance},
  author={Vissers, Michael R and Hubmayr, Johannes and Sandberg, Martin and Chaudhuri, Saptarshi and Bockstiegel, Clint and Gao, Jiansong},
  journal={Appl. Phys. Lett.},
  volume={107},
  number={6},
  year={2015},
  publisher={AIP Publishing}
}

@article{Kirsh2021,
  title={Linear and nonlinear properties of a compact high-kinetic-inductance WSi multimode resonator},
  author={Kirsh, Naftali and Svetitsky, Elisha and Goldstein, Samuel and Pardo, Guy and Hachmo, Ori and Katz, Nadav},
  journal={Phys. Rev. A},
  volume={16},
  number={4},
  pages={044017},
  year={2021},
  publisher={APS}
}

@article{Zhao2020,
  title={Nonlinear properties of supercurrent-carrying single-and multi-layer thin-film superconductors},
  author={Zhao, Songyuan and Withington, Stafford and Goldie, David J and Thomas, Chris N},
  journal={J. Low Temp. Phys.},
  volume={199},
  number={1},
  pages={34--44},
  year={2020},
  publisher={Springer}
}

@article{bottcher2024circuit,
  title={Circuit quantum electrodynamics detection of induced two-fold anisotropic pairing in a hybrid superconductor--ferromagnet bilayer},
  author={B{\o}ttcher, CGL and Poniatowski, NR and Grankin, A and Wesson, ME and Yan, Z and Vool, U and Galitski, VM and Yacoby, A},
  journal={Nature Physics},
  pages={1--7},
  year={2024},
  publisher={Nature Publishing Group UK London}
}

@article{weitzel2023sharpness,
  title={Sharpness of the Berezinskii-Kosterlitz-Thouless transition in disordered NbN films},
  author={Weitzel, Alexander and Pfaffinger, Lea and Maccari, Ilaria and Kronfeldner, Klaus and Huber, Thomas and Fuchs, Lorenz and Mallord, James and Linzen, Sven and Il’ichev, Evgeni and Paradiso, Nicola and others},
  journal={Physical Review Letters},
  volume={131},
  number={18},
  pages={186002},
  year={2023},
  publisher={APS}
}

@article{phan2022detecting,
  title={Detecting induced p$\pm$ip pairing at the Al-InAs interface with a quantum microwave circuit},
  author={Phan, D and Senior, J and Ghazaryan, Areg and Hatefipour, M and Strickland, WM and Shabani, J and Serbyn, Maksym and Higginbotham, Andrew P},
  journal={Physical Review Letters},
  volume={128},
  number={10},
  pages={107701},
  year={2022},
  publisher={APS}
}

@article{schmidt2018ballistic,
  title={A ballistic graphene superconducting microwave circuit},
  author={Schmidt, Felix E and Jenkins, Mark D and Watanabe, Kenji and Taniguchi, Takashi and Steele, Gary A},
  journal={Nature communications},
  volume={9},
  number={1},
  pages={4069},
  year={2018},
  publisher={Nature Publishing Group UK London}
}

@article{giaever1960electron,
  title={Electron tunneling between two superconductors},
  author={Giaever, Ivar},
  journal={Physical Review Letters},
  volume={5},
  number={10},
  pages={464},
  year={1960},
  publisher={APS}
}

@article{gittleman1965nonlinear,
  title={Nonlinear reactance of superconducting films},
  author={Gittleman, J and Rosenblum, B and Seidel, TE and Wicklund, AW},
  journal={Physical Review},
  volume={137},
  number={2A},
  pages={A527},
  year={1965},
  publisher={APS}
}

@article{kubo2020superfluid,
  title={Superfluid flow in disordered superconductors with Dynes pair-breaking scattering: Depairing current, kinetic inductance, and superheating field},
  author={Kubo, Takayuki},
  journal={Physical Review Research},
  volume={2},
  number={3},
  pages={033203},
  year={2020},
  publisher={APS}
}

@article{enpuku1993modulation,
  title={Modulation of kinetic inductance of high Tc superconducting thin films with bias current},
  author={Enpuku, Keiji and Hoashi, Masakazu and Doi, Hideki Doi Hideki and Kisu, Takanobu Kisu Takanobu},
  journal={Japanese journal of applied physics},
  volume={32},
  number={9R},
  pages={3804},
  year={1993},
  publisher={IOP Publishing}
}

@article{anthore2003density,
  title={Density of states in a superconductor carrying a supercurrent},
  author={Anthore, Anne and Pothier, Hugues and Esteve, Daniel},
  journal={Physical review letters},
  volume={90},
  number={12},
  pages={127001},
  year={2003},
  publisher={APS}
}

@article{ho2012wideband,
  title={A wideband, low-noise superconducting amplifier with high dynamic range},
  author={Ho Eom, Byeong and Day, Peter K and LeDuc, Henry G and Zmuidzinas, Jonas},
  journal={Nature Physics},
  volume={8},
  number={8},
  pages={623--627},
  year={2012},
  publisher={Nature Publishing Group UK London}
}

@article{luomahaara2014kinetic,
  title={Kinetic inductance magnetometer},
  author={Luomahaara, Juho and Vesterinen, Visa and Gr{\"o}nberg, Leif and Hassel, Juha},
  journal={Nature communications},
  volume={5},
  number={1},
  pages={4872},
  year={2014},
  publisher={Nature Publishing Group UK London}
}

@article{Lee2005,
  title={Doping-dependent nonlinear Meissner effect and spontaneous currents in high-$T_{\mathrm{C}}$ superconductors},
  author={Lee, Sheng-Chiang and Sullivan, Mathew and Ruchti, Gregory R and Anlage, Steven M and Palmer, Benjamin S and Maiorov, B and Osquiguil, E},
  journal={Phys. Rev. B},
  volume={71},
  number={1},
  pages={014507},
  year={2005},
  publisher={APS}
}

@article{Wilcox2022,
  title={Observation of the non-linear Meissner effect},
  author={Wilcox, JA and Grant, MJ and Malone, L and Putzke, C and Kaczorowski, D and Wolf, T and Hardy, F and Meingast, C and Analytis, JG and Chu, J-H and others},
  journal={Nat. Commun.},
  volume={13},
  number={1},
  pages={1201},
  year={2022},
  publisher={Nature Publishing Group UK London}
}

@article{Makita2022,
   author = {J. Makita and C. Sundahl and G. Ciovati and C. B. Eom and A. Gurevich},
   doi = {10.1103/PhysRevResearch.4.013156},
   issn = {26431564},
   issue = {1},
   journal = {Phys. Rev. R},
  
   publisher = {American Physical Society},
   title = {Nonlinear Meissner effect in $\mathrm{Nb_{3}Sn}$ coplanar resonators},
   volume = {4},
   year = {2022},
}

@article{carrington1999absence,
  title={Absence of nonlinear Meissner effect in $\mathrm{YBa_{2}Cu_{3}O_{6.95}}$},
  author={Carrington, A and Giannetta, RW and Kim, JT and Giapintzakis, J},
  journal={Phys. Rev. B},
  volume={59},
  number={22},
  pages={R14173},
  year={1999},
  publisher={APS}
}

@article{khvalyuk2024near,
  title={Near power-law temperature dependence of the superfluid stiffness in strongly disordered superconductors},
  author={Khvalyuk, Anton V and Charpentier, Thibault and Roch, Nicolas and Sac{\'e}p{\'e}, Benjamin and Feigel'Man, Mikhail V},
  journal={Phys. Rev. B},
  volume={109},
  number={14},
  pages={144501},
  year={2024},
  publisher={APS}
}

@article{ghosal2001inhomogeneous,
  title={Inhomogeneous pairing in highly disordered s-wave superconductors},
  author={Ghosal, Amit and Randeria, Mohit and Trivedi, Nandini},
  journal={Phys. Rev. B},
  volume={65},
  number={1},
  pages={014501},
  year={2001},
  publisher={APS}
}

@article{bretheau2017tunnelling,
  title={Tunnelling spectroscopy of Andreev states in graphene},
  author={Bretheau, Landry and Wang, {Joel I-Jan} and Pisoni, Riccardo and Watanabe, Kenji and Taniguchi, Takashi and {Jarillo-Herrero, Pablo}},
  journal={Nat. Phys.},
  volume={13},
  number={8},
  pages={756--760},
  year={2017},
  publisher={Nature Publishing Group UK London}
}

@article{teknowijoyo2018nodeless,
  title={Nodeless superconductivity in the type-II Dirac semimetal PdTe 2: London penetration depth and pairing-symmetry analysis},
  author={Teknowijoyo, Serafim and Jo, Na Hyun and Scheurer, Mathias S and Tanatar, Makariy A and Cho, Kyuil and Bud'ko, Sergey L and Orth, Peter P and Canfield, Paul C and Prozorov, Ruslan},
  journal={Physical Review B},
  volume={98},
  number={2},
  pages={024508},
  year={2018},
  publisher={APS}
}

@article{schwinger1951gauge,
  title={On gauge invariance and vacuum polarization},
  author={Schwinger, Julian},
  journal={Physical Review},
  volume={82},
  number={5},
  pages={664},
  year={1951},
  publisher={APS}
}

@article{berdyugin2022out,
  title={schwinger},
  author={Berdyugin, Alexey I and Xin, Na and Gao, Haoyang and Slizovskiy, Sergey and Dong, Zhiyu and Bhattacharjee, Shubhadeep and Kumaravadivel, Piranavan and Xu, Shuigang and Ponomarenko, LA and Holwill, Matthew and others},
  journal={Science},
  volume={375},
  number={6579},
  pages={430--433},
  year={2022},
  publisher={American Association for the Advancement of Science}
}

@article{Dahm1999,
  title={Nonlinear current response of a d-wave superfluid},
  author={Dahm, Thomas and Scalapino, DJ},
  journal={Phys. Rev. B},
  volume={60},
  number={18},
  pages={13125},
  year={1999},
  publisher={APS}
}

@article{yip1992nonlinear,
  title={Nonlinear Meissner effect in CuO superconductors},
  author={Yip, SK and Sauls, JA},
  journal={Phys. Rev. Lett.},
  volume={69},
  number={15},
  pages={2264},
  year={1992},
  publisher={APS}
}

@article{dahm1996theory,
  title={Theory of microwave intermodulation in a high-$T_{\mathrm{C}}$ superconducting microstrip resonator},
  author={Dahm, Thomas and Scalapino, DJ},
  journal={Appl. Phys. Lett.},
  volume={69},
  number={27},
  pages={4248--4250},
  year={1996},
  publisher={American Institute of Physics}
}

@article{li1998nonlinear,
  title={Is the nonlinear Meissner effect unobservable?},
  author={Li, M-R and Hirschfeld, PJ and W{\"o}lfle, P},
  journal={Phys. Rev. Lett.},
  volume={81},
  number={25},
  pages={5640},
  year={1998},
  publisher={APS}
}

@article{Cao2018,
   author = {Yuan Cao and Valla Fatemi and Shiang Fang and Kenji Watanabe and Takashi Taniguchi and Efthimios Kaxiras and Pablo Jarillo-Herrero},
   doi = {10.1038/nature26160},
   %issn = {14764687},
   %issue = {7699},
   journal = {Nature},
   pages = {43-50},
   pmid = {29512651},
   publisher = {Nature Publishing Group},
   title = {Unconventional superconductivity in magic-angle graphene superlattices},
   volume = {556},
   year = {2018},
}

@article{cao2021nematicity,
  title={Nematicity and competing orders in superconducting magic-angle graphene},
  author={Cao, Yuan and Rodan-Legrain, Daniel and Park, Jeong Min and Yuan, Noah FQ and Watanabe, Kenji and Taniguchi, Takashi and Fernandes, Rafael M and Fu, Liang and Jarillo-Herrero, Pablo},
  journal={Science},
  volume={372},
  number={6539},
  pages={264--271},
  year={2021},
  publisher={American Association for the Advancement of Science}
}

@article{peotta2015superfluidity,
  title={Superfluidity in topologically nontrivial flat bands},
  author={Peotta, Sebastiano and T{\"o}rm{\"a}, P{\"a}ivi},
  journal={Nat. Commun.},
  volume={6},
  number={1},
  pages={8944},
  year={2015},
  publisher={Nature Publishing Group UK London}
}

@article{xie2020topology,
  title={Topology-bounded superfluid weight in twisted bilayer graphene},
  author={Xie, Fang and Song, Zhida and Lian, Biao and Bernevig, B Andrei},
  journal={Phys. Rev. Lett.},
  volume={124},
  number={16},
  pages={167002},
  year={2020},
  publisher={APS}
}

@article{torma2022superconductivity,
  title={Superconductivity, superfluidity and quantum geometry in twisted multilayer systems},
  author={T{\"o}rm{\"a}, P{\"a}ivi and Peotta, Sebastiano and Bernevig, Bogdan A},
  journal={Nat. Rev. Phys.},
  volume={4},
  number={8},
  pages={528--542},
  year={2022},
  publisher={Nature Publishing Group UK London}
}

@article{hu2019geometric,
  title={Geometric and conventional contribution to the superfluid weight in twisted bilayer graphene},
  author={Hu, Xiang and Hyart, Timo and Pikulin, Dmitry I and Rossi, Enrico},
  journal={Phys. Rev. Lett.},
  volume={123},
  number={23},
  pages={237002},
  year={2019},
  publisher={APS}
}

@article{wu2019topological,
  title={Topological chiral superconductivity with spontaneous vortices and supercurrent in twisted bilayer graphene},
  author={Wu, Fengcheng},
  journal={Phys. Rev. B},
  volume={99},
  number={19},
  pages={195114},
  year={2019},
  publisher={APS}
}

@article{julku2020superfluid,
  title={Superfluid weight and Berezinskii-Kosterlitz-Thouless transition temperature of twisted bilayer graphene},
  author={Julku, Aleksi and Peltonen, Teemu J and Liang, Long and Heikkil{\"a}, Tero T and T{\"o}rm{\"a}, P{\"a}ivi},
  journal={Phys. Rev. B},
  volume={101},
  number={6},
  pages={060505},
  year={2020},
  publisher={APS}
}

@article{tian2023evidence,
  title={Evidence for Dirac flat band superconductivity enabled by quantum geometry},
  author={Tian, Haidong and Gao, Xueshi and Zhang, Yuxin and Che, Shi and Xu, Tianyi and Cheung, Patrick and Watanabe, Kenji and Taniguchi, Takashi and Randeria, Mohit and Zhang, Fan and others},
  journal={Nature},
  volume={614},
  number={7948},
  pages={440--444},
  year={2023},
  publisher={Nature Publishing Group UK London}
}

@article{Tormarev2023,
  title = {Essay: Where Can Quantum Geometry Lead Us?},
  author = {T\"orm\"a, P\"aivi},
  journal = {Phys. Rev. Lett.},
  volume = {131},
 % issue = {24},
  pages = {240001},
  numpages = {7},
  year = {2023},
 
  publisher = {American Physical Society},
  doi = {10.1103/PhysRevLett.131.240001},
}

@article{hofmann2022heuristic,
  title={Heuristic bounds on superconductivity and how to exceed them},
  author={Hofmann, Johannes S and Chowdhury, Debanjan and Kivelson, Steven A and Berg, Erez},
  journal={npj quantum materials},
  volume={7},
  number={1},
  pages={83},
  year={2022},
  publisher={Nature Publishing Group UK London}
}

@incollection{kosterlitz1973ordering,
  title={Ordering, metastability and phase transitions in two-dimensional systems},
  author={Kosterlitz, John Michael and Thouless, David James},
  booktitle={Basic Notions Of Condensed Matter Physics},
  pages={493--515},
  year={1973},
  publisher={CRC Press}
}

@article{emery1995importance,
  title={Importance of phase fluctuations in superconductors with small superfluid density},
  author={Emery, VJ and Kivelson, SA},
  journal={Nature},
  volume={374},
  number={6521},
  pages={434--437},
  year={1995},
  publisher={Nature Publishing Group UK London}
}

@article{cao2020strange,
  title={Strange metal in magic-angle graphene with near Planckian dissipation},
  author={Cao, Yuan and Chowdhury, Debanjan and Rodan-Legrain, Daniel and Rubies-Bigorda, Oriol and Watanabe, Kenji and Taniguchi, Takashi and Senthil, T and Jarillo-Herrero, Pablo},
  journal={Phys. Rev. Lett.},
  volume={124},
  number={7},
  pages={076801},
  year={2020},
  publisher={APS}
}

@article{cao2018correlated,
  title={Correlated insulator behaviour at half-filling in magic-angle graphene superlattices},
  author={Cao, Yuan and Fatemi, Valla and Demir, Ahmet and Fang, Shiang and Tomarken, Spencer L and Luo, Jason Y and Sanchez-Yamagishi, Javier D and Watanabe, Kenji and Taniguchi, Takashi and Kaxiras, Efthimios and others},
  journal={Nature},
  volume={556},
  number={7699},
  pages={80--84},
  year={2018},
  publisher={Nature Publishing Group UK London}
}

@article{hao2021electric,
  title={Electric field--tunable superconductivity in alternating-twist magic-angle trilayer graphene},
  author={Hao, Zeyu and Zimmerman, AM and Ledwith, Patrick and Khalaf, Eslam and Najafabadi, Danial Haie and Watanabe, Kenji and Taniguchi, Takashi and Vishwanath, Ashvin and Kim, Philip},
  journal={Science},
  volume={371},
  number={6534},
  pages={1133--1138},
  year={2021},
  publisher={American Association for the Advancement of Science}
}

@article{park2021tunable,
  title={Tunable strongly coupled superconductivity in magic-angle twisted trilayer graphene},
  author={Park, Jeong Min and Cao, Yuan and Watanabe, Kenji and Taniguchi, Takashi and Jarillo-Herrero, Pablo},
  journal={Nature},
  volume={590},
  number={7845},
  pages={249--255},
  year={2021},
  publisher={Nature Publishing Group UK London}
}

@article{park2022robust,
  title={Robust superconductivity in magic-angle multilayer graphene family},
  author={Park, Jeong Min and Cao, Yuan and Xia, Li-Qiao and Sun, Shuwen and Watanabe, Kenji and Taniguchi, Takashi and Jarillo-Herrero, Pablo},
  journal={Nat. Mater.},
  volume={21},
  number={8},
  pages={877--883},
  year={2022},
  publisher={Nature Publishing Group UK London}
}

@article{zhang2022promotion,
  title={Promotion of superconductivity in magic-angle graphene multilayers},
  author={Zhang, Yiran and Polski, Robert and Lewandowski, Cyprian and Thomson, Alex and Peng, Yang and Choi, Youngjoon and Kim, Hyunjin and Watanabe, Kenji and Taniguchi, Takashi and Alicea, Jason and others},
  journal={Science},
  volume={377},
  number={6614},
  pages={1538--1543},
  year={2022},
  publisher={American Association for the Advancement of Science}
}

@article{burg2022emergence,
  title={Emergence of correlations in alternating twist quadrilayer graphene},
  author={Burg, G William and Khalaf, Eslam and Wang, Yimeng and Watanabe, Kenji and Taniguchi, Takashi and Tutuc, Emanuel},
  journal={Nat. Mater.},
  volume={21},
  number={8},
  pages={884--889},
  year={2022},
  publisher={Nature Publishing Group UK London}
}

@article{andrei2020graphene,
  title={Graphene bilayers with a twist},
  author={Andrei, Eva Y and MacDonald, Allan H},
  journal={Nat. Mater.},
  volume={19},
  number={12},
  pages={1265--1275},
  year={2020},
  publisher={Nature Publishing Group UK London}
}

@article{prozorov2006magnetic,
  title={Magnetic penetration depth in unconventional superconductors},
  author={Prozorov, Ruslan and Giannetta, Russell W},
  journal={Supercond. Sci. Technol.},
  volume={19},
  number={8},
  pages={R41},
  year={2006},
  publisher={IOP Publishing}
}

@book{hardy2002magnetic,
  title={Magnetic penetration depths in cuprates: A short review of measurement techniques and results},
  bookTitle={The Gap Symmetry and Fluctuations in High-Tc Superconductors},
  author={Hardy, WN and Kamal, S and Bonn, DA},
  year={2002},
  publisher={Springer US}
}

@book{tinkham2004introduction,
  title={Introduction to Superconductivity: Second Edition},
  author = {Tinkham, Michael},
publisher = {Dover Publications},
}

@article{oh2021evidence,
  title={Evidence for unconventional superconductivity in twisted bilayer graphene},
  author={Oh, Myungchul and Nuckolls, Kevin P and Wong, Dillon and Lee, Ryan L and Liu, Xiaomeng and Watanabe, Kenji and Taniguchi, Takashi and Yazdani, Ali},
  journal={Nature},
  volume={600},
  number={7888},
  pages={240--245},
  year={2021},
  publisher={Nature Publishing Group UK London}
}

@article{mishra2009lifting,
  title = {Lifting of nodes by disorder in extended-$s$-state superconductors: Application to ferropnictides},
  author = {Mishra, V. and Boyd, G. R. and Graser, S. and Maier, T. and Hirschfeld, P. J. and Scalapino, D. J.},
  journal = {Phys. Rev. B},
  volume = {79},
  issue = {9},
  pages = {094512},
  numpages = {9},
  year = {2009},

  publisher = {American Physical Society},
  doi = {10.1103/PhysRevB.79.094512},
}

@article{cho2016energy,
  title={Energy gap evolution across the superconductivity dome in single crystals of (Ba1- x K x) Fe2As2},
  author={Cho, Kyuil and Ko{\'n}czykowski, Marcin and Teknowijoyo, Serafim and Tanatar, Makariy A and Liu, Yong and Lograsso, Thomas A and Straszheim, Warren E and Mishra, Vivek and Maiti, Saurabh and Hirschfeld, Peter J and others},
  journal={Science Advances},
  volume={2},
  number={9},
  pages={e1600807},
  year={2016},
  publisher={American Association for the Advancement of Science}
}

@article{Raychaudhuri_2022,
doi = {10.1088/1361-648X/ac360b},
%url = {https://dx.doi.org/10.1088/1361-648X/ac360b},
year = {2021},

publisher = {IOP Publishing},
volume = {34},
number = {8},
pages = {083001},
author = {Pratap Raychaudhuri and Surajit Dutta},
title = {Phase fluctuations in conventional superconductors},
journal = {Journal of Physics: Condensed Matter},}

@article{jarjour2023superfluid,
  title={Superfluid response of an atomically thin gate-tuned van der Waals superconductor},
  author={Jarjour, Alexander and Ferguson, GM and Schaefer, Brian T and Lee, Menyoung and Loh, Yen Lee and Trivedi, Nandini and Nowack, Katja C},
  journal={Nat. Commun.},
  volume={14},
  number={1},
  pages={2055},
  year={2023},
  publisher={Nature Publishing Group UK London}
}

@article{yong2013robustness,
  title={Robustness of the Berezinskii-Kosterlitz-Thouless transition in ultrathin NbN films near the superconductor-insulator transition},
  author={Yong, Jie and Lemberger, TR and Benfatto, L and Ilin, K and Siegel, M},
  journal={Phys. Rev. B},
  volume={87},
  number={18},
  pages={184505},
  year={2013},
  publisher={APS}
}

@article{ma2020moire,
  title={Moir{\'e} band topology in twisted bilayer graphene},
  author={Ma, Chao and Wang, Qiyue and Mills, Scott and Chen, Xiaolong and Deng, Bingchen and Yuan, Shaofan and Li, Cheng and Watanabe, Kenji and Taniguchi, Takashi and Du, Xu and others},
  journal={Nano letters},
  volume={20},
  number={8},
  pages={6076--6083},
  year={2020},
  publisher={ACS Publications}
}

@article{wang2013one,
  title={One-dimensional electrical contact to a two-dimensional material},
  author={Wang, Lei and Meric, I and Huang, PY and Gao, Q and Gao, Y and Tran, H and Taniguchi, T and Watanabe, Kenji and Campos, LM and Muller, DA and others},
  journal={Science},
  volume={342},
  number={6158},
  pages={614--617},
  year={2013},
  publisher={American Association for the Advancement of Science}
}

@article{penttilä2024flatband,
    title={Flat-band ratio and quantum metric in the superconductivity of modified Lieb lattices}, 
    author={Reko P. S. Penttilä and Kukka-Emilia Huhtinen and Päivi Törmä},
    year={2024},
    journal = {Preprint at arXiv},
    url ={https://arxiv.org/abs/2404.12993}
}

@article{yan2016flux,
  title={The flux qubit revisited to enhance coherence and reproducibility},
  author={Yan, Fei and Gustavsson, Simon and Kamal, Archana and Birenbaum, Jeffrey and Sears, Adam P and Hover, David and Gudmundsen, Ted J and Rosenberg, Danna and Samach, Gabriel and Weber, Steven and others},
  journal={Nat. Commun.},
  volume={7},
  number={1},
  pages={1--9},
  year={2016},
  publisher={Nature Publishing Group}
}

@article{wang2019coherent,
  title={Coherent control of a hybrid superconducting circuit made with graphene-based van der Waals heterostructures},
  author={Wang, Joel I-Jan and Rodan-Legrain, Daniel and Bretheau, Landry and Campbell, Daniel L and Kannan, Bharath and Kim, David and Kjaergaard, Morten and Krantz, Philip and Samach, Gabriel O and others},
  journal={ Nat. Nanotechnol.},
  volume={14},
  number={2},
  pages={120--125},
  year={2019},
  publisher={Nature Publishing Group}
}

@Article{Probst2015,
  author    = {S. Probst and F. B. Song and P. A. Bushev and A. V. Ustinov and M. Weides},
  title     = {Efficient and robust analysis of complex scattering data under noise in microwave resonators},
  journal   = {Rev. Sci. Instrum.},
  year      = {2015},
  volume    = {86},
  number    = {2},
  pages     = {024706},
 
  doi       = {10.1063/1.4907935},
  publisher = {{AIP} Publishing},
}

\section*{Acknowledgements}

We acknowledge helpful discussions with Päivi Törmä, Leonid Levitov, Senthil Todadri, Masaki Roppongi, Kota Ishihara, Taisei Kitamura, and Yoichi Yanase. The authors thank Lamia Ateshian, David Rower, Patrick Harrington, and the device packaging team at MIT Lincoln Laboratory for technical assistance.

This research was funded in part by the US Army Research Office grant no. W911NF-22-1-0023, by the National Science Foundation QII-TAQS grant no. OMA-1936263, by the Air Force Office of Scientific Research grant no. FA2386-21-1-4058, and by the Under Secretary of Defense for Research and Engineering under Air Force Contract No. FA8702-15-D-0001. 
P.J-H. acknowledges support by the National Science Foundation (DMR-1809802), the Gordon and Betty Moore Foundation’s EPiQS Initiative through Grant No. GBMF9463, the Fundacion Ramon Areces, and the CIFAR Quantum Materials program. 
D. R-L. acknowledges support from the Rafael del Pino Foundation. K.W. and T.T. acknowledge support from the JSPS KAKENHI (Grant Numbers 21H05233 and 23H02052) and World Premier International Research Center Initiative (WPI), MEXT, Japan.
Any opinions, findings, conclusions or recommendations expressed in this material are those of the author(s) and do not necessarily reflect the views of the Under Secretary
of Defense for Research and Engineering or the U.S. Government.

\section*{Author Contributions}
J.\^I-j.W. conceived and designed the experiment. M.T., J.\^I-j.W., T.H.D., and M.H. performed the microwave simulation. M.T., J.\^I-j.W., T.H.D., S.Z., D. R-L., D.K.K., B.M.N., K.S., M.E.S. contributed to the device fabrication. M.T., J.\^I-j.W, T.H.D., S.Z., D. R-L., A. A., and B.K. participated in the measurements. M.T., J.\^I-j.W., and M.H. analyzed the data. K.W. and T.T. grew the hBN crystal. J.\^I-j.W., M.T., and W.D.O. led the paper writing, and all other authors contributed to the text. J.A.G., T.P.O., S.G., P.J-H., J.\^I-j.W., and W.D.O supervised the project.   
\section*{Competing Interests Statement}
The authors declare no competing interests.

\section*{Correspondence and requests for materials}
should be addressed to Joel \^I-j. Wang, Pablo Jarillo-Herrero, or William D. Oliver.

\section*{Extended data}

\begin{extendedfigure}[H]
    \centering
    \includegraphics[width=0.8\textwidth]{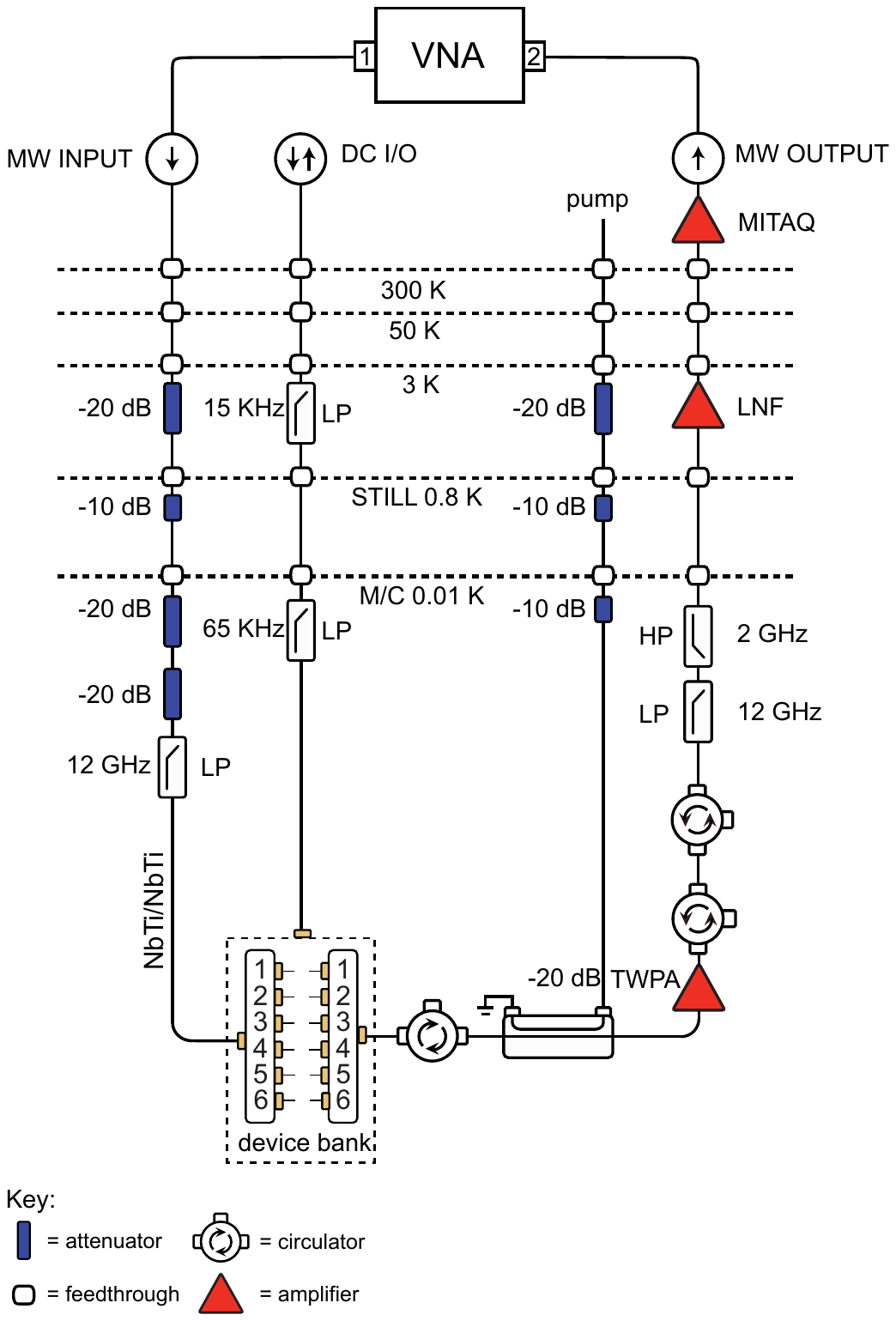}
    \caption{\textbf{Wiring diagram of the microwave and DC characterization setup.} }
    \label{fig:wiring}
\end{extendedfigure}

\begin{extendedfigure}[H]
    \centering
    \includegraphics[width=1\textwidth]{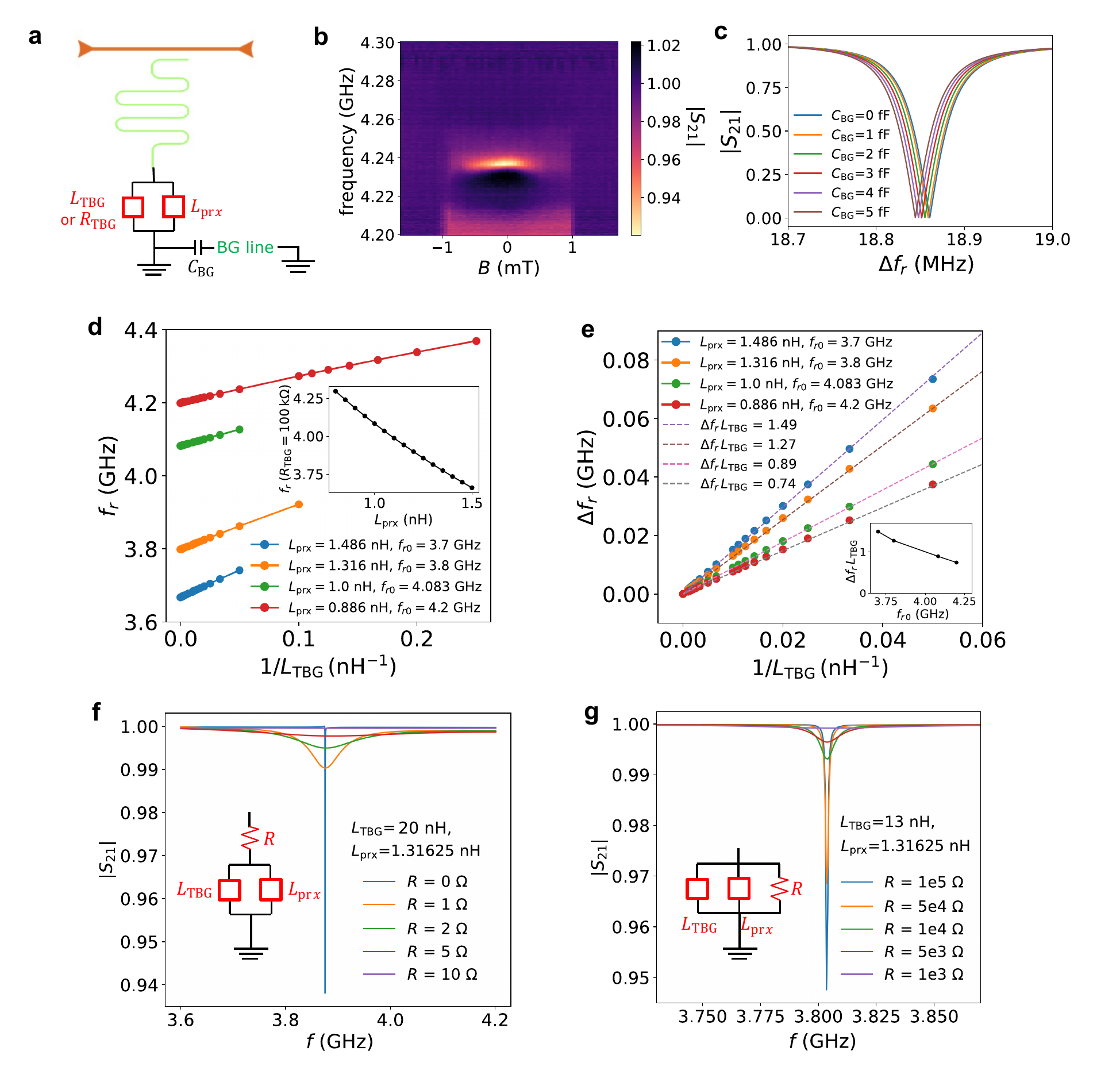}
    \caption{\textbf{Model and simulation of the resonator with termination.} \textbf{a,} Model used for the microwave simulation, including the through line, resonator, termination, and backgate. The termination includes both the MATBG (resistance or inductance, depending on bias point) and the proximitized edge inductance (independent of bias point). See main text. \textbf{b,} Out-of-plane magnetic field dependence of the resonance of Bernal-stacked bilayer graphene biased near its charge neutrality point. \textbf{c,} Simulated $|S_{21}|$ as a function of $\Delta f_{\mathrm{r}}=f_{\mathrm{r}}-3.8$ GHz for different values of $C_{\mathrm{BG}}$ of the model. \textbf{d,} Simulated resonance frequency as a function of $1/L_{\mathrm{TBG}}$ for different values of $L_{\mathrm{prx}}$. $C_{\mathrm{BG}}$ is fixed  at 3 fF. \textbf{e,} Data in panel \textbf{d,}, with the baseline resonance frequency subtracted. Dashed lines are the linear fit to the simulation data. Inset depicts the slope of the linear fit as a function of offset frequency $f_{\mathrm{r0}}$, which depends on $L_{\mathrm{prx}}$.
    \textbf{f, g,} Simulations of possible series (\textbf{f}) and parallel (\textbf{g}) contact resistances between the MATBG and the resonator, ground, and leads. Our results are most consistent with superconducting contacts, i.e., zero series resistance and large parallel resistance.}
    \label{fig:simulation}
\end{extendedfigure}

\begin{extendedfigure}[H]
    \centering
    \includegraphics[width=0.75\textwidth]{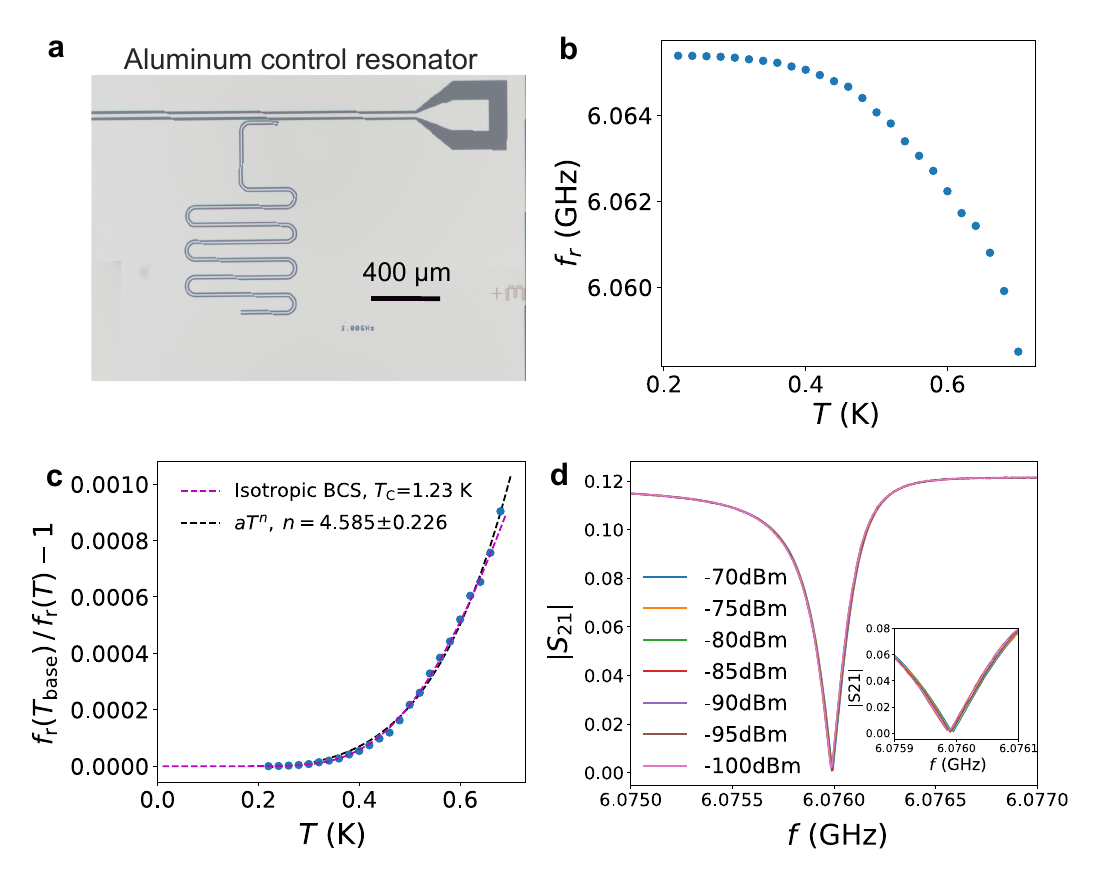}
    \caption{\textbf{Temperature and power dependence of the resonant frequency for the aluminum-only $\lambda /4$ resonator (control resonator).} 
    \textbf{a,} Microscope image of the aluminum-only $\lambda/4$ resonator, terminated directly to ground with Al. 
    \textbf{b,} Temperature dependence of the resonance for the aluminum-only $\lambda/4$ resonator. 
    \textbf{c,} Temperature dependence of $f_{\mathrm{r}}(T_{\mathrm{base}})\,/\,f_{\mathrm{r}}(T)-1$ for the aluminum-only $\lambda/4$ resonator as fit with the isotropic BCS model (exponential fit, purple dashed line) and a power-law fit (black dashed line). 
    \textbf{d,} $|S_{21}|$ for the aluminum-only $\lambda/4$ resonator at $T_{\mathrm{base}}$=20 mK for different microwave powers. The power independence indicates the negligible kinetic inductance for the thick (250 nm) aluminum resonator and termination.
    }
    \label{fig:cont}
\end{extendedfigure}

\begin{extendedfigure}[H]
    \centering
    \includegraphics[width=0.75\textwidth]{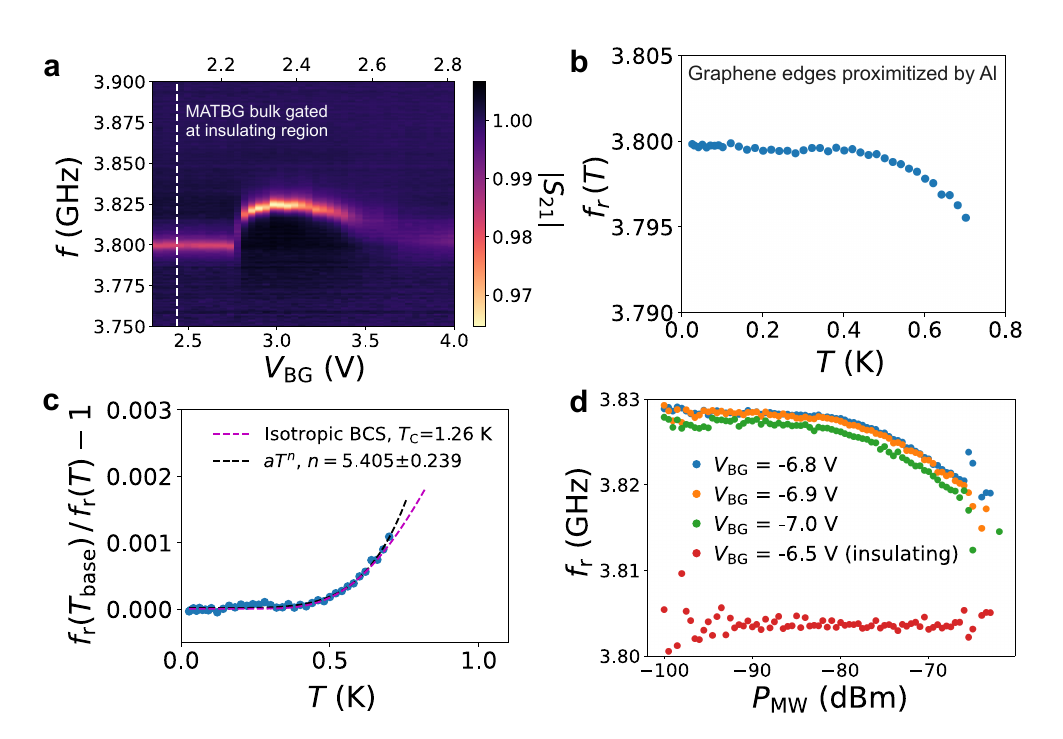}
    \caption{\textbf{Temperature and power dependence of the resonant frequency for the resonator terminated by the Al-prozimitized graphene edge.} 
    \textbf{a,} Gate dependence of the resonance in the MATBG-terminated resonator. The insulating region is indicated by a vertical white dashed line $V_{\mathrm{BG}}$ =2.44 V. The frequency at this point is used as the reference point when determining a frequency shift. 
    \textbf{b,} Temperature dependence of the resonance for the MATBG-terminated resonator in the insulating region $V_{\mathrm{BG}}$=2.44 V. 
    \textbf{c,} Temperature dependence of $f_{\mathrm{r}}(T_{\mathrm{base}})\,/\,f_{\mathrm{r}}(T)-1$ for the MATBG-terminated resonator in the insulating region $V_{\mathrm{BG}}$=2.44 V and fit with the isotropic BCS model (purple dashed line) and the power-law fit (black dashed line). The resonance no longer shifts below about 0.5 K.
    \textbf{d,} Microwave power dependence of $f_{\mathrm{r}}$ at $V_{\mathrm{BG}}$=-6.8, -6.9, -7.0 V (where the MATBG is superconducting) and $V_{\mathrm{BG}}$=-6.5 V (where the MATBG is insulating). In contrast to the MATBG traces, the proximitized edge does not vary substantially with power, indicating that its kinetic inductance is adopted from the aluminum leads and is relatively small compared with the bulk MATBG.}
    \label{fig:insu}
\end{extendedfigure}

\section*{Data availability}
The data that support the findings of this study are available from the corresponding author upon reasonable request and with the cognizance of our US Government sponsors who funded the work.

\newpage
\part*{Supplementary Information}

\section{Lorentzian fitting of the resonance}
\begin{figure}[H]
    \centering
    \includegraphics[width=0.7\textwidth]{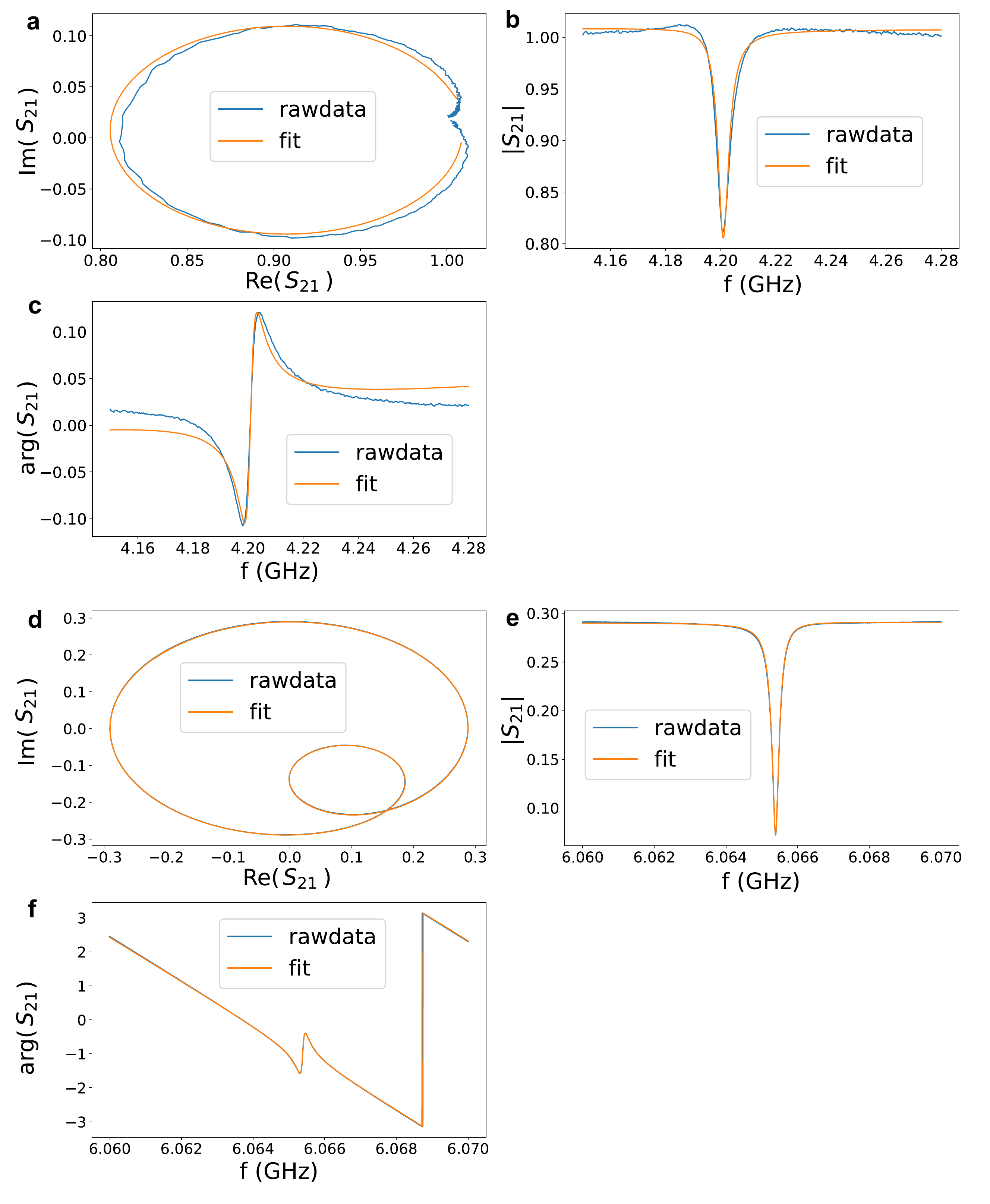}
    \caption{\textbf{Lorentzian fitting of the resonance.} 
    \textbf{a, and d,} Raw data (blue)  
    and Lorentzian fit (orange) of $S_{21}$ signal plotted in the complex plane for the MATBG resonator (a) and the control resonator (d). 
    \textbf{b, and e,} Raw data and Lorentzian fit of the $S_{21}$ magnitude for the MATBG resonator (b) and the control resonator (e). 
    \textbf{c, and f,} Raw data and Lorentzian fit of the $S_{21}$ phase for the MATBG resonator (c) and the control resonator (f).}
    \label{fig:lorentz}
\end{figure}

Fig.~\ref{fig:lorentz} shows an example of the Lorentzian fitting procedure for a resonance of the MATBG resonator and the control resonator. We use the fitting function:
\begin{align}\label{eq:lorentz}
S_{21}(f) = a e^{-\alpha-2\pi ifd}\;\Bigg[1-\frac{(Q_{l}/Q_{c})e^{i\phi}}{1+2iQ_{l}(f-f_{\mathrm{r}})/f_{\mathrm{r}}}\Bigg],
\end{align}
where $a$ is a scaling coefficient determined by loss, attenuation, and amplification of the measurement line, $\alpha$ is the phase offset of the signal, $d$ is the length of the measurement line, and $f_{r}$ is the resonant frequency. The loaded quality factor is given by $Q_{l}=(Q_{i}^{-1}+Q_{c}^{-1})^{-1}$, where $Q_{i}$ and $Q_{c}$ are the internal and coupling quality factors, respectively.
By fitting $S_{21}$ in the complex plane, we extract the following parameters for the MATBG resonator: $Q_{i}=1044.7\pm28.5$, $Q_{c}=4161\pm54.4$, $f_{\mathrm{r}}=4.2008154\pm0.0000737$ GHz and the control resonator: $Q_{i}=74833.2\pm1830.2$, $Q_{c}=24926.9\pm60.3$, $f_{\mathrm{r}}=6.0653899\pm0.0000011$ GHz.

\section{Estimation of Fermi velocity}\label{section:vf}

In systems with a low inelastic scattering rate, the velocity of electron flow can reach the Fermi velocity, leading to non-equilibrium carrier generation, known as the Schwinger mechanism. Since MATBG has a Dirac cone with a much lower Fermi velocity than monolayer graphene, this effect occurs at relatively small currents, resulting in a critical-current-like behavior in the normal state\cite{tian2023evidence, berdyugin2022out, schwinger1951gauge}.  

\begin{figure}[H]
    \centering
    \includegraphics[width=0.9\textwidth]{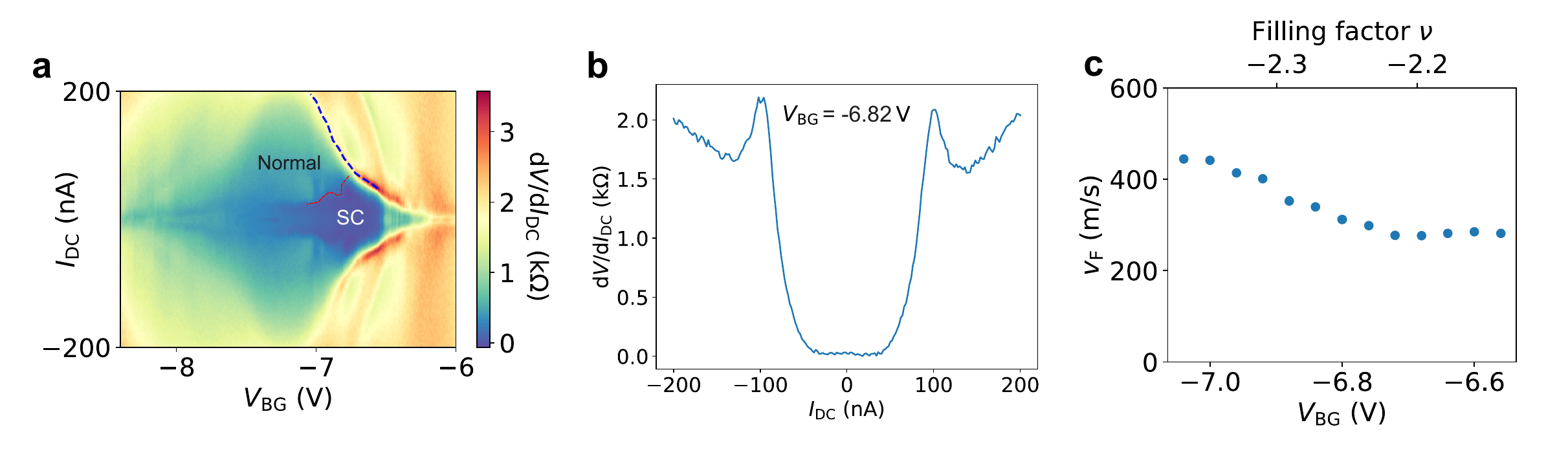}
    \caption{\textbf{Estimation of Fermi velocity.} 
    \textbf{a,} Differential resistance $dV/dI_{\mathrm{DC}}$ as a function of DC bias current $I_{\mathrm{DC}}$ and gate voltage $V_{\mathrm{BG}}$. The blue dashed line indicates $I_{cn}$.
    \textbf{b,} Differential resistance $dV/dI_{\mathrm{DC}}$ as a function of DC bias current $I_{\mathrm{DC}}$ at $V_{\mathrm{BG}}=-6.82$ V.
    \textbf{c,} Gate dependence of the Fermi velocity, extracted from $v_{n}=J_{cn}/\tilde{n}e$.
    }
    \label{fig:vf}
\end{figure}

Here, the critical current in the normal state, $I_{cn}$, is determined by the Fermi velocity as $v_{n} = \frac{J_{cn}}{\tilde{n}e}$, where $J_{cn}$ is the critical current density and $\tilde{n}$ is the effective carrier density measured relative to $\mid\nu\mid$=2, providing an estimate of the Fermi velocity. In a previous study under a finite magnetic field, which was used to suppress superconductivity, it was reported that $I_{cn}$ has a ‘bell-like’ feature in vicinity of $\nu$=2. The superconducting critical current $I_{\mathrm{C}}$ coincident with $I_{cn}$ at the underdoped region but split in the overdoped region~\cite{tian2023evidence}. We observe a similar behavior in the peak of $\frac{dV}{dI}$ (Fig. \ref{fig:vf}a) and estimate the Fermi velocity from the larger peak of $\frac{dV}{dI}$ (Fig. \ref{fig:vf}c). The conventional contribution to the superfluid density in Figure 3 is calculated based on this Fermi velocity.

\section{Determination of \texorpdfstring{$T_{\mathrm{C}}$}{Tc}}\label{section:Tc}
\begin{figure}[H]
    \centering
    \includegraphics[width=0.6\textwidth]{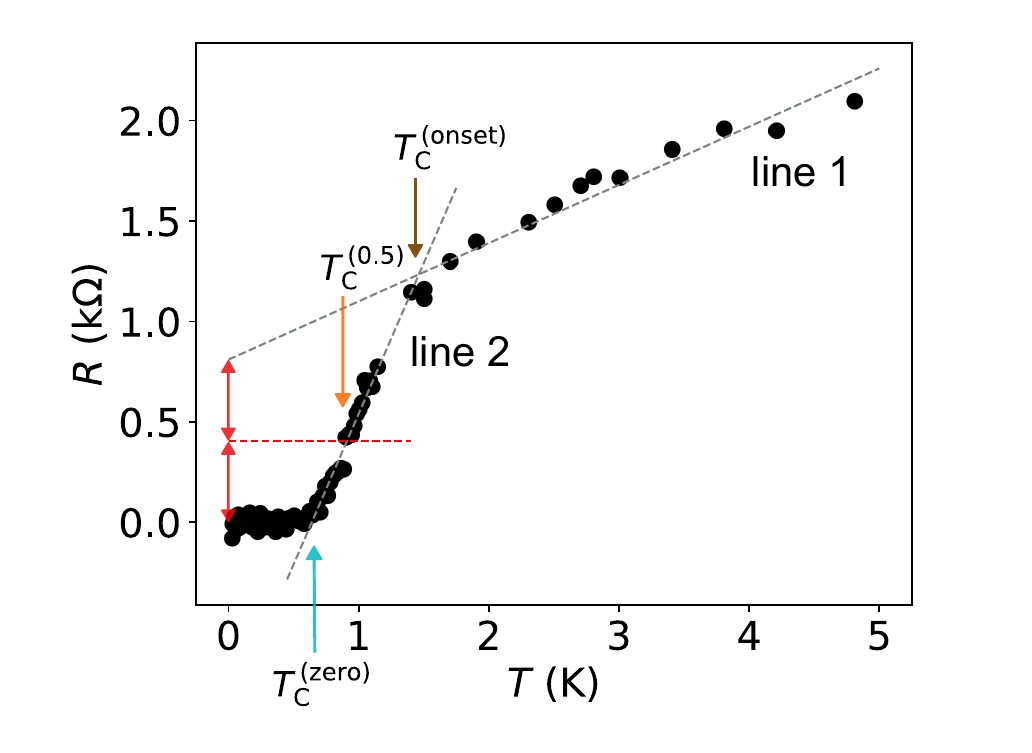}
    \caption{\textbf{Determination of $T_{\mathrm{C}}$.} The temperature dependence of the MATBG resistance at $V_{\mathrm{BG}} = -6.7$ V. Gray dashed lines represent linear fits. Red arrows and a red dashed line indicate the half value of line 1 at zero temperature. Critical temperatures are marked by arrows of different colors.}
    \label{fig:Tc}
\end{figure}

The critical temperatures in Figure 3 in the main text are determined as follows  (see also Fig. S3):

\begin{enumerate} 
\item{Linear fits are performed for the normal region and the region where the resistance drops (gray dashed lines in Fig.~\ref{fig:Tc}), referred to as line 1, and line 2. }

\item{$T_{\mathrm{C}}^{\mathrm{(onset)}}$ is determined by the intersection of line 1 and line 2.}

\item{$T_{\mathrm{C}}^{\mathrm{(0.5)}}$ is determined by the point where line 2 crosses half the value of the intercept of line 1.}

\item{$T_{\mathrm{C}}^{\mathrm{(zero)}}$ is determined by the point where line 2 crosses zero resistance.}
\end{enumerate}

\section{Power-law fitting at different gate voltages}\label{section:power}

\begin{figure}[H]
    \centering
    \includegraphics[width=1\textwidth]{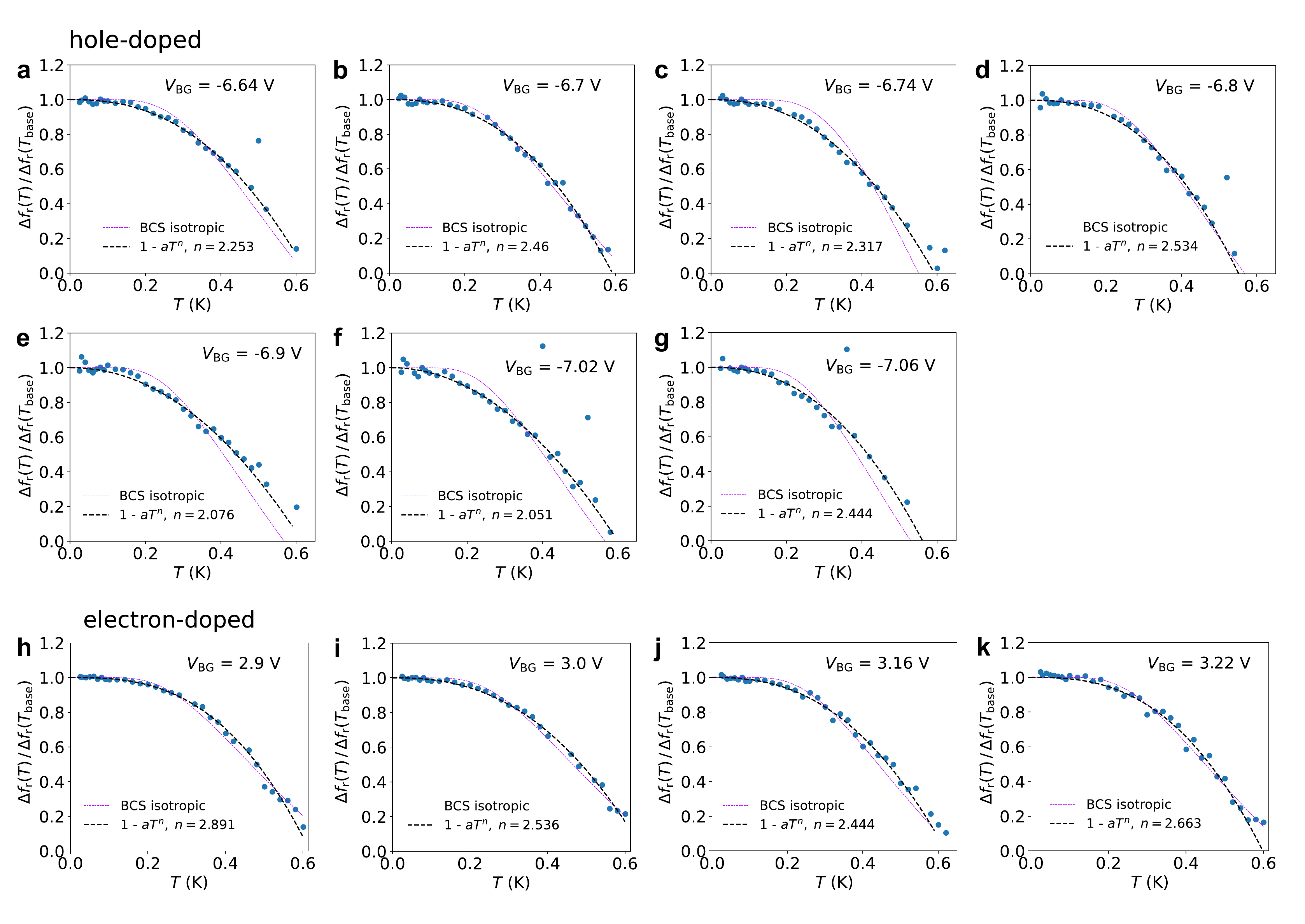}
    \caption{\textbf{Power-law fitting at various gate voltages.} The temperature dependence of $\Delta f_{\mathrm{r}}(T)/\Delta f_{\mathrm{r}}(T_{\mathrm{base}})$ for $T < T_{\mathrm{BKT}}$ in the hole-doped (\textbf{a-g}) and electron-doped (\textbf{h-k}) regimes. Blue dots represent the experimental data, black dashed lines depict the power-law fits, and green dashed lines represent the exponential function in the BCS isotropic model.}
    \label{fig:power}
\end{figure}

\section{Fitting the temperature dependence with an extended s-wave model for conventional superfluid stiffness}\label{section:fitting}
The temperature dependence of the superfluid stiffness within single-band Fermi liquid theory $D_{\mathrm{s}}^{\mathrm{(conv)}}(T)$ is formulated using the Mattis-Bardeen equation as~\autocite{prozorov2006magnetic, hardy2002magnetic, tinkham2004introduction}:
\begin{align}\label{eq:MB}
\frac{D_{\mathrm{s}}^{\mathrm{(conv)}}(T)}{D_{\mathrm{s}}^{\mathrm{(conv)}}(0)} = 1 + \frac{1}{\pi}\int_{0}^{2\pi}d\phi\int_{\Delta(T,\phi)}^{\infty}\frac{\partial f}{\partial E}\frac{E}{\sqrt{E^{2}-\Delta(T,\phi)^{2}}}dE,
\end{align}
where $f$ is Fermi distribution function and $\phi$ is the argument in momentum space.

To model a superconducting gap with a continuous degree of anisotropy, we assume the following angle-dependent superconducting gap:
\begin{align}\label{eq:s-wave}
    \Delta(T,\phi) = \Delta_{0}(T)g(\phi)=\Delta_{0}(T)\frac{1+\alpha\cos{(2\phi)}}{\sqrt{1+\alpha^{2}/2}},
\end{align}
where $\alpha=0$ corresponds to an isotropic gap, and $\alpha=1$ is a nodal gap. The factor 2 in $\cos{(2\phi)}$, which represents two-fold rotational symmetry of the superconducting gap, is chosen without loss of generality of the rotational symmetry, because the resulting temperature dependence $D_{\mathrm{s}}^{\mathrm{(conv)}}(T)$ does not depend on this factor.

Within a weak-coupling BCS theory, $\Delta_{0}(T)$ in Eq.\ref{eq:s-wave} is determined from the following self-consistent equation:
\begin{align}\label{eq:self}
\int_{0}^{\infty}d\epsilon \int_{0}^{2\pi}d\phi\biggl[\frac{\tanh{(\frac{\epsilon^{2}+\Delta_{0}^{2}(T)g^{2}(\phi)}{2T})}}{\sqrt{\epsilon^{2}+\Delta_{0}^{2}(T)g^{2}(\phi)}} 
- \frac{\tanh({\frac{\epsilon}{2T_{\mathrm{C}}}})}{\epsilon}\biggr]g^{2}(\phi) = 0.
\end{align}

\begin{figure}[H]
    \centering
    \includegraphics[width=0.84\textwidth]{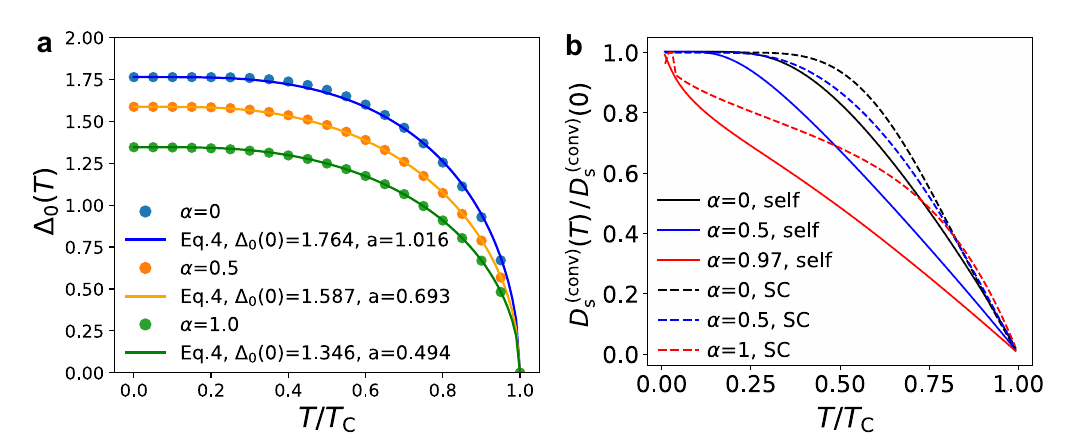}
    \caption{\textbf{Calculation of the temperature dependence of $D_{\mathrm{s}}^{\mathrm{(conv)}}$ based on the Mattis-Bardeen formula.} 
    \textbf{a,} Self-consistently calculated $\Delta_{0}(T)$ for different anisotropy parameters $\alpha$. Dots are the numerical results from the self-consistent equation, and lines are a fit using Eq.~\ref{eq:aprox}. 
    \textbf{b,} Temperature dependence of $D_{s}$ for different parameters of the formula. Black, blue, and red indicate anisotropy parameter $\alpha=$0, 0.5, 1, respectively. For the solid lines (self), $\Delta_{0}(0)$ and $a$ in Eq.~\ref{eq:aprox} are determined self-consistently using Eq.~\ref{eq:self}. For the dashed lines (SC, strong coupling), they are fixed at $\Delta_{0}(0)=3k_{\mathrm{B}}T_{\mathrm{C}}$ and $a=1$.}
    \label{fig:self}
\end{figure}

Fig.~\ref{fig:self}a presents solutions of this equation for several anisotropy parameters $\alpha$. 
It is known that the solutions of the self-consistent gap equation can be approximated by the analytical formula: 
\begin{align}\label{eq:aprox}
\Delta_{0}(T) = \Delta_{0}(0)\tanh{\biggl[\frac{\pi T}{\Delta_{0}(0)}\sqrt{a(T_{\mathrm{C}}/T -1)}\biggr]},
\end{align}
with two free parameters, $a$ and $\Delta_{0}(0)$ (Fig.~\ref{fig:self}a).
Substituting Eq.\ref{eq:aprox} into Eq.\ref{eq:s-wave} and Eq.\ref{eq:s-wave} into Eq.\ref{eq:MB} yields $\frac{D_{\mathrm{s}}^{\mathrm{(conv)}}(T)}{D_{\mathrm{s}}^{\mathrm{(conv)}}(0)}$ (solid lines in Fig.~\ref{fig:self}b). 
In scanning tunneling microscope studies, Andreev reflection spectroscopy suggest a strong-coupling $\Delta_{0}(0)/k_B T_{\mathrm{C}}\simeq3$ in MATBG~\autocite{oh2021evidence}. Using this value and assuming $a=1$ in Eq.\ref{eq:s-wave}, we obtain $\frac{D_{\mathrm{s}}^{\mathrm{(conv)}}(T)}{D_{\mathrm{s}}^{\mathrm{(conv)}}(0)}$ (dashed lines in Fig.~\ref{fig:self}b).

\begin{figure}[H]
    \centering
    \includegraphics[width=0.8\textwidth]{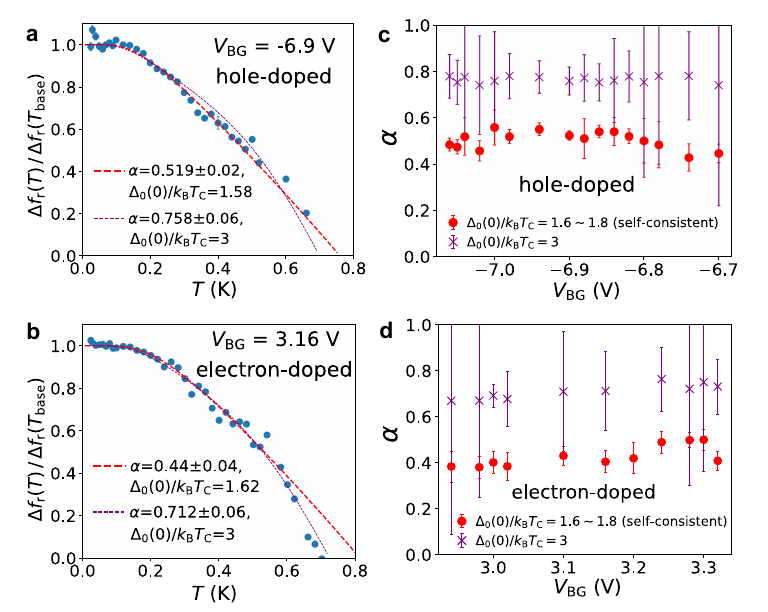}
    \caption{\textbf{Analysis of the temperature dependence based on the Mattis-Bardeen formula.} %
    \textbf{a, and b,} The temperature dependence of $\Delta f_{\mathrm{r}}(T)\,/\,\Delta f_{\mathrm{r}}(T_{\mathrm{base}})$ and fit functions based on the Mattis-Bardeen formula in the hole-doped (\textbf{a}) and the electron-doped (\textbf{b}) regimes. The red dashed curves represent best-fits to an extended s-wave model using anisotropy parameters $\alpha$ and $T_{\mathrm{C}}$ as fitting parameters, where $\Delta_{0}(0)/k_{\mathrm{B}}T_{\mathrm{C}}$ is determined self-consistently. The purple dashed curves represent fittings with the fixed value of $\Delta_{0}(0)/k_{\mathrm{B}}T_{\mathrm{C}}=3$. %
    \textbf{c,and d,} Backgate dependence of the anisotropy paramemter $\alpha$ obtained from fitting. Red and purple data points fit with the self-consistently determined $\Delta_{0}(0)/k_{\mathrm{B}}T_{\mathrm{C}}$ and the $\Delta_{0}(0)/k_{\mathrm{B}}T_{\mathrm{C}}=3$, respectively.}
    \label{fig:temp}
\end{figure}

Using this procedure for both the weak- and strong-coupling cases, we fit $\Delta f_{\mathrm{r}}(T)\,/\,\Delta f_{\mathrm{r}}(T_{\mathrm{base}})$ over a temperature range from the refrigerator base temperature to the Berezinskii-Kosterlitz-Thouless transition temperature $T_{\mathrm{BKT}}$, determined by $8e^{2}k_{\mathrm{B}}T_{\mathrm{BKT}}/\pi \hbar^{2}=D_{\mathrm{s}}(T_{\mathrm{BKT}})$ (see section~\ref{section:BKT}), up to which this formulation is valid~\autocite{Raychaudhuri_2022, jarjour2023superfluid, yong2013robustness}. 
The results at $V_{\mathrm{BG}}=-6.9$ V and $V_{\mathrm{BG}}=3.16$V are shown with red and purple dashed lines in Figs.~\ref{fig:temp}a and~\ref{fig:temp}b, respectively, yielding anisotropy parameters $\alpha$ = 0.52 (weak), 0.76 (strong) for hole-doped MATBG and 0.44 (weak), 0.71 (strong) for electron-doped MATBG. By performing the same fitting procedure over the superconducting dome, we obtain $\alpha=0.4\sim 0.8$ for both the hole-doped and electron-doped regimes (Figs.~\ref{fig:temp}c-d). Critical temperatures obtained from this fitting are comparable with $T_{\mathrm{C}}$ values determined from DC resistance mesurements. \\

\section{Berezinskii-Kosterlitz-Thouless transition and universal jumps}\label{section:BKT}
The Berezinskii-Kosterlitz-Thouless (BKT) theory of thermally excited vortex–antivortex pairs in 2D superconductors predicts a super-to-normal phase transition marked by a discontinuous drop in superfluid stiffness at a temperature $T_{\mathrm{BKT}}$ determined by the formula $8e^{2}k_{\mathrm{B}}T_{\mathrm{BKT}}/\pi \hbar^{2}=D_{\mathrm{s}}(T_{\mathrm{BKT}})$. 
The temperature dependence of the conventional superfluid stiffness below $T_{\mathrm{BKT}}$ follows the Mattis-Bardeen formula within the framework of Fermi liquid theory, and it exhibits an abrupt reduction at $T_{\mathrm{BKT}}$, which is called a ``universal jump''~\autocite{Raychaudhuri_2022, jarjour2023superfluid, yong2013robustness}. Therefore, power-law fitting in the main text and Mattis-Bardeen fitting in the previous section are done over the temperature range $T<T_{\mathrm{BKT}}$. 

\begin{figure}[H]
    \centering
    \includegraphics[width=0.9\textwidth]{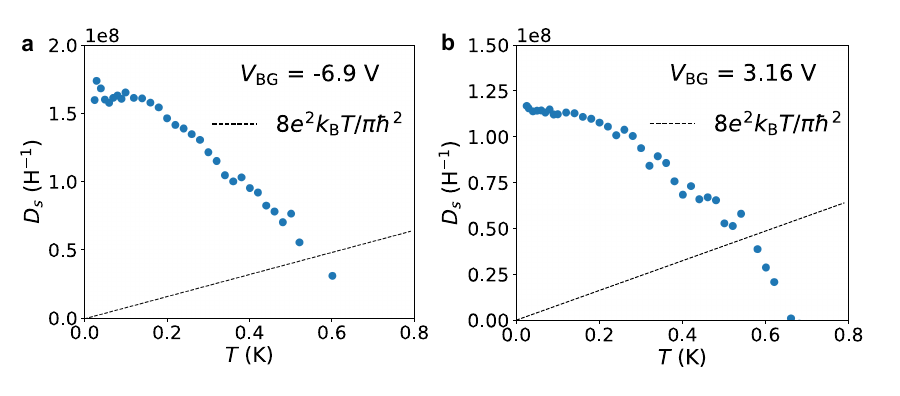}
    \caption{\textbf{Temperature dependence of the superfluid stiffness and determination of the BKT temperature.} \textbf{a, and b,} Temperature dependence of $D_{\mathrm{s}}$ in hole-doped (a) and electron-doped (b) regimes, respectively. The intersection of the black dasehd line and the data determine $T_{\mathrm{BKT}}$.}
    \label{fig:univ}
\end{figure}

Fig.~\ref{fig:univ} shows the temperature dependence of the measured $D_{\mathrm{s}}$ and $8e^{2}k_{\mathrm{B}}T/\pi \hbar^{2}$, where $T_{\mathrm{BKT}}$ is determined by their intersection. 

\section{Additional bias-dependence data}\label{section:Ebias}
\begin{figure}[H]
    \centering
    \includegraphics[width=0.5\textwidth]{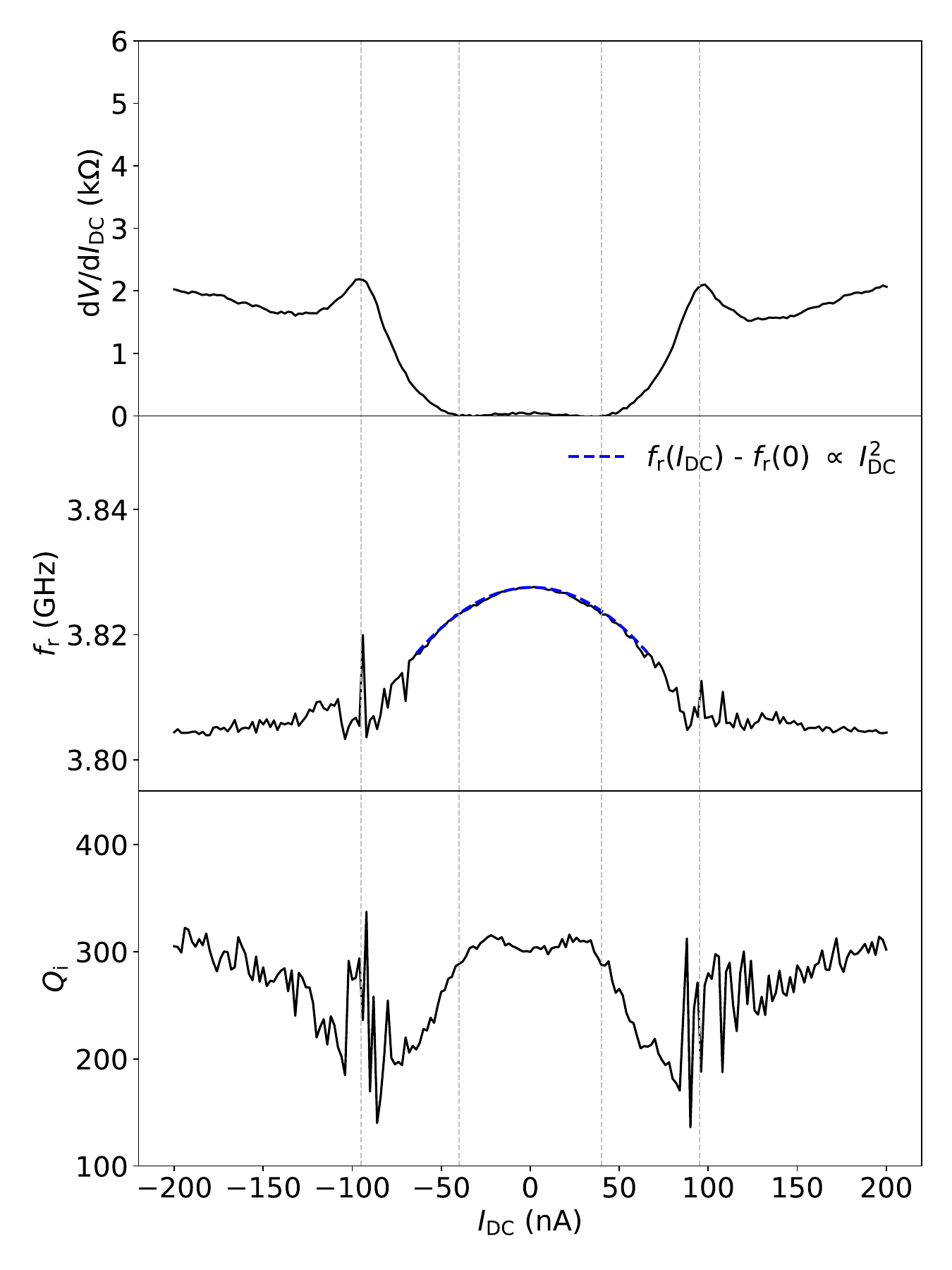}
    \caption{\textbf{Bias current dependence of $dV/dI_{\mathrm{DC}}$, $f_{\mathrm{r}}$, and internal Q factor $Q_{\mathrm{i}}$ at $V_{\mathrm{BG}}$ = -6.8 V.} The blue dashed line in the center panel is quadratic fit.}
    \label{fig:biasQ}
\end{figure}

Fig.~\ref{fig:biasQ} displays the bias current dependence of $dV/dI_{\mathrm{DC}}$, $f_{\mathrm{r}}$, and the internal quality factor $Q_{\mathrm{i}}$ at $V_{\mathrm{BG}}$ = -6.7 V. The top and center panels are the same with those in Fig. 5 
of the main text. Notably, $Q_{\mathrm{i}}$ remains nearly constant at low bias currents (below 45 nA), similar to the behavior of $dV/dI_{\mathrm{DC}}$. This confirms that no significant normal regions develop in the MATBG. The change in $D_{s}$ is mainly due to a spatially uniform reduction in the superfluid density. 

In the intermediate region ($I_{\mathrm{DC}}$ = 50$\sim$100 nA), dV/dI is finite while quadratic behavior of $f_{r}$ persist. We can think of two possible reasons:
\begin{enumerate}
    \item The superfluid density is suppressed for various reasons, such as Cooper pair depairing or unbound vortex-antivortex pairs in a 2D superconductor. As a result, the current path is no longer uniformly superconducting, with some normal regions contributing to finite series resistance, evident in the increasing dV/dI and decreasing Qi. The quadratic behavior in this regime can be attributed to the remaining superfluid, which is expected to exhibit similar behavior to that in the low-bias regime ($I_{\mathrm{DC}} \leq 45 $ nA).
    \item Our device could have different path for DC and microwave current, therefore DC path might be partially normal while MW path remains superconducting in this regime. However, this is unlikely a dominating factor as the the consistency between dV/dI and $Q_{i}$ suggest otherwise.
\end{enumerate}

Additionally, the small finite differential resistance at zero bias current could be attributed to 1) one or more small non-superconducting regions along the DC current path between the superconducting electrodes. These regions may result from twist-angle inhomogeneity, impurities, or other forms of disorder, leading to enhanced resistance at zero bias~\autocite{giaever1960electron}; and 2) a mismatch between the DC current path and the microwave current path due to the device configuration.

\begin{figure}[H]
    \centering
    \includegraphics[width=0.65\textwidth]{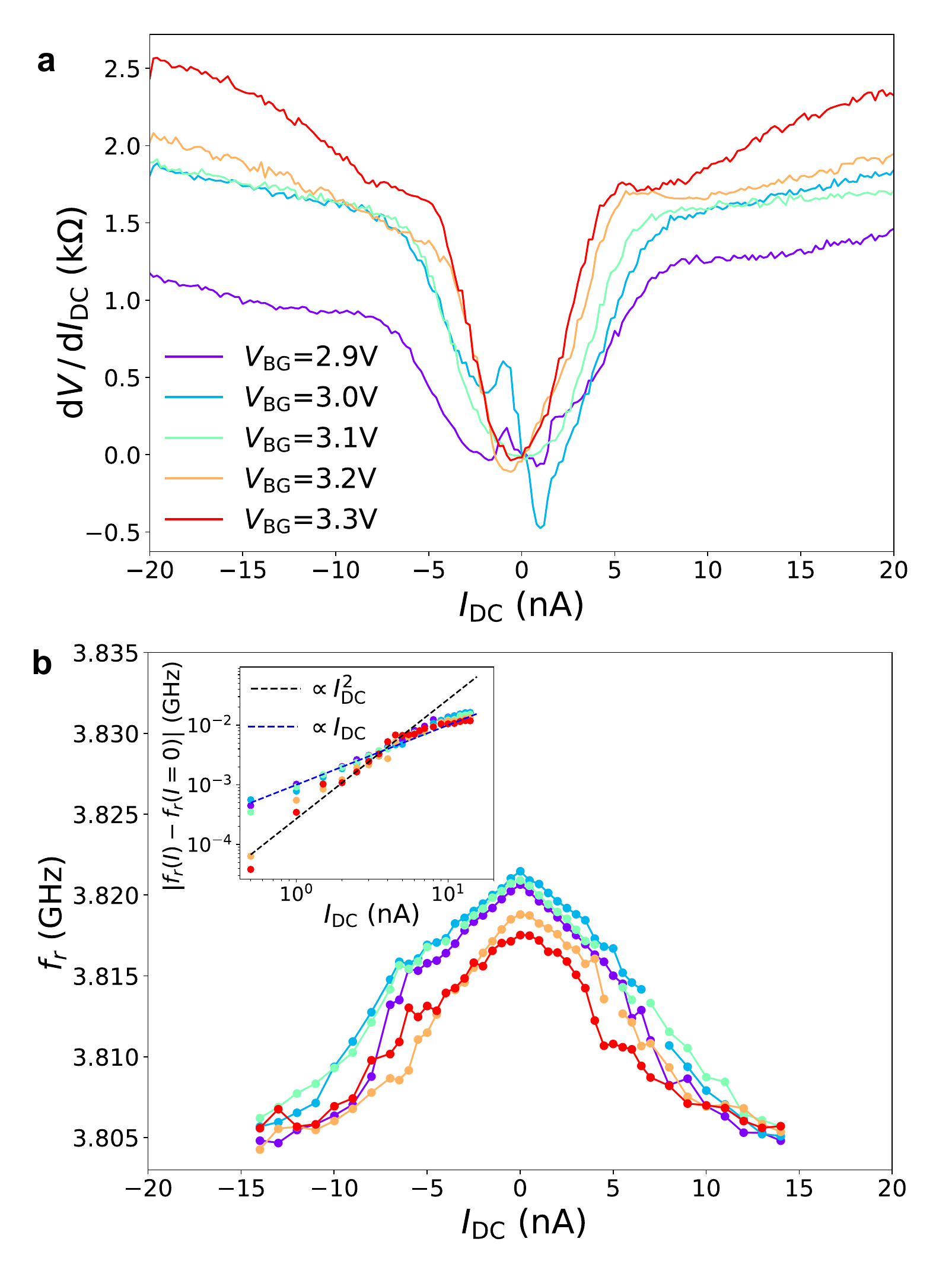}
    \caption{\textbf{Bias current dependence of the resistance and the resonant frequency in the electron-doped regime.} 
    \textbf{a,} DC resistance $R$ as a function of DC bias current $I_{\mathrm{DC}}$ for different $V_{\mathrm{BG}}$. 
    \textbf{b,} Resonant frequency $f_{\mathrm{r}}$ as a function of $I_{\mathrm{DC}}$. Inset: Log-log plot of $f_{\mathrm{r}}(I)-f_{\mathrm{r}}(0)$ versus $I_{\mathrm{DC}}$. The blue and black dashed line indicate linear and quadratic dependence, respectively.}
    \label{fig:Ebias}
\end{figure}

In the electron-doped regime, the critical current is much smaller, likely due to a smaller spatial extent of the superconducting state (a larger effective aspect ratio leads to a larger effective current density in the superconducting region), see Fig.~\ref{fig:Ebias}a. The bias dependence of the resonant frequency is linear instead of quadratic over the superconducting dome, as shown in  Fig.\ref{fig:Ebias}b. This bias dependence in the electron-doped regime likely indicates that the superconducting region comprises islands connected via narrow channels, where finite bias currents can readily break superconductivity in the narrow channels; consequently, the averaged superfluid stiffness rapidly decreases.

The nonlinear Meissner effect (NLME) has been discussed as an origin of the linear reduction in superfluid stiffness as a function of magnetic field or bias current in nodal superconductors~\autocite{Dahm1999, yip1992nonlinear, dahm1996theory, Lee2005, Wilcox2022}. 
Considering the NLME in a disordered system, a crossover from quadratic to linear dependence is theoretically expected as the current increases, and the crossover current is proportional to the degree of disorder~\autocite{Dahm1999}. In our device, we observe a linear dependence only in the electron-doped region, which, based on the smaller critical current, is affected more strongly by disorder than the hole-doped regime. This trend is opposite to the expected disorder dependence~\autocite{Dahm1999}, suggesting that the observed linear dependence in the electron-doped region is not due to the NLME. 
In addition, we note that the absence of the NLME has also been reported in the nodal superconductor YBCO~\cite{carrington1999absence, li1998nonlinear}. Furthermore, the existence of weak links can cause a linear dependence extrinsically in nodeless materials~\autocite{Makita2022}, indicating that a linear dependence or the presence of NLME is not a necessary nor a sufficient condition of nodal superconductivity. 

\section{Additional sample}\label{section:B3ST5}

\begin{figure}[H]
    \centering
    \includegraphics[width=0.65\textwidth]{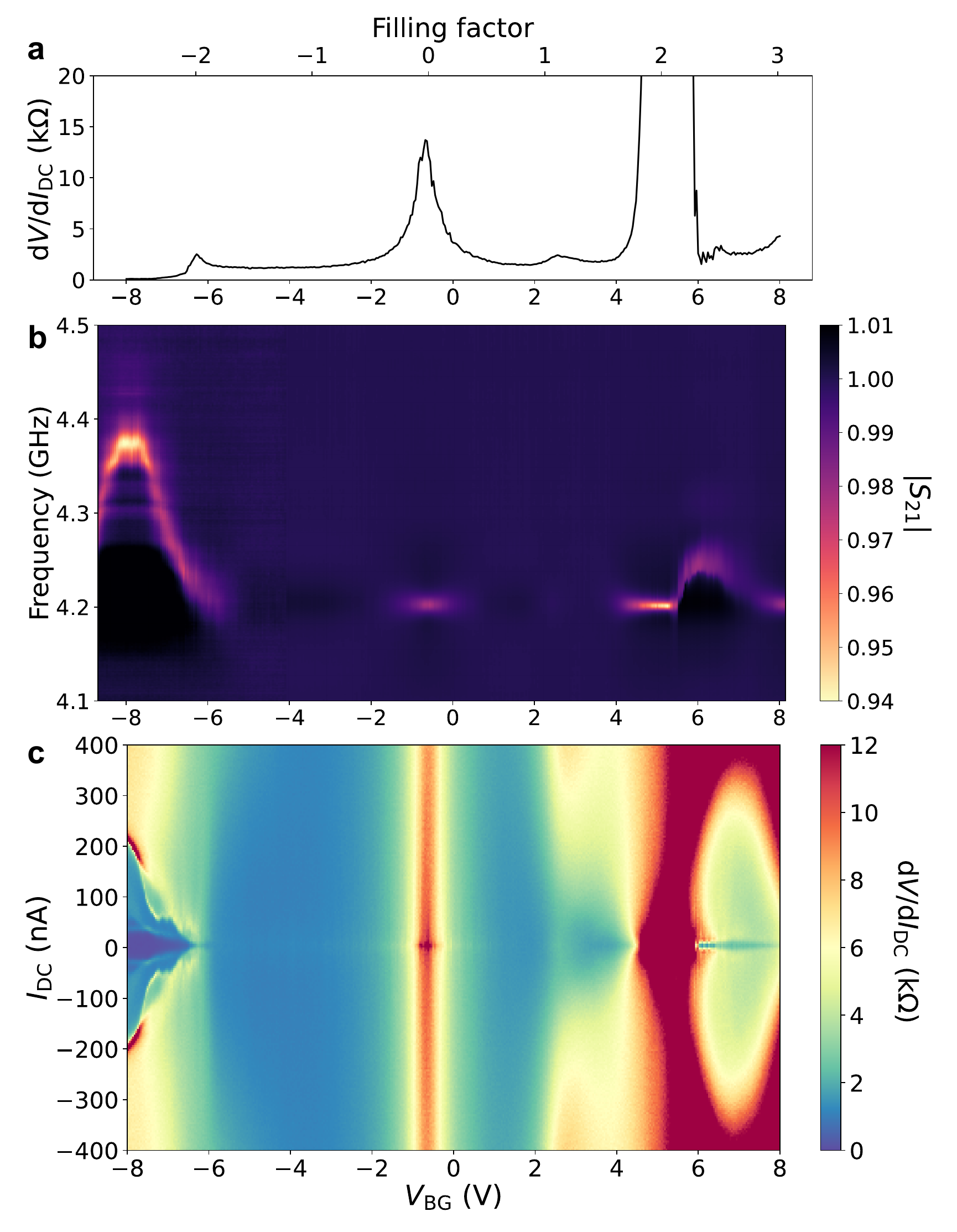}
    \caption{\textbf{Gate dependence of DC and microwave response in the second device.} \textbf{a,} DC resistance $R$ as a function of the backgate voltage $V_{\mathrm{BG}}$. Top axis represents the filling factor $\nu$.
    \textbf{b,} Microwave transmission coefficient $|S_{21}|$ versus $V_{\mathrm{BG}}$. The resonant frequency (bright line) shifts within the zero-resistance region in panel (a), near filling factors $\nu=\pm 2$. The resonance remains essentially constant within the high resistance region. 
    \textbf{c,} Differential resistance $\mathrm{d}V/\mathrm{d}I_{\mathrm{DC}}$ as a function of $V_{\mathrm{BG}}$ and DC bias current $I_{\mathrm{DC}}$.}
    \label{fig:B3ST5gate}
\end{figure}

In another device with the same design and fabrication process, we observed qualitatively similar results, although it exhibited more spatial inhomogeneity, which results in multiple $T_{\mathrm{C}}$ and $I_{\mathrm{C}}$ values.
Fig.~\ref{fig:B3ST5gate} shows the DC and microwave response of this second device as a function of $V_{\mathrm{BG}}$. 
The behavior qualitatively matches the first device with the following notable differences: 
1. the electron-doped superconducting region does not exhibit complete zero-resistance whereas the first device shows zero resistance. 
2. the hole-doped superconducting region shows multiple critical currents, which is thought to be due to the spatial inhomogeneity of the twist angle.

\begin{figure}[H]
    \centering
    \includegraphics[width=0.85\textwidth]{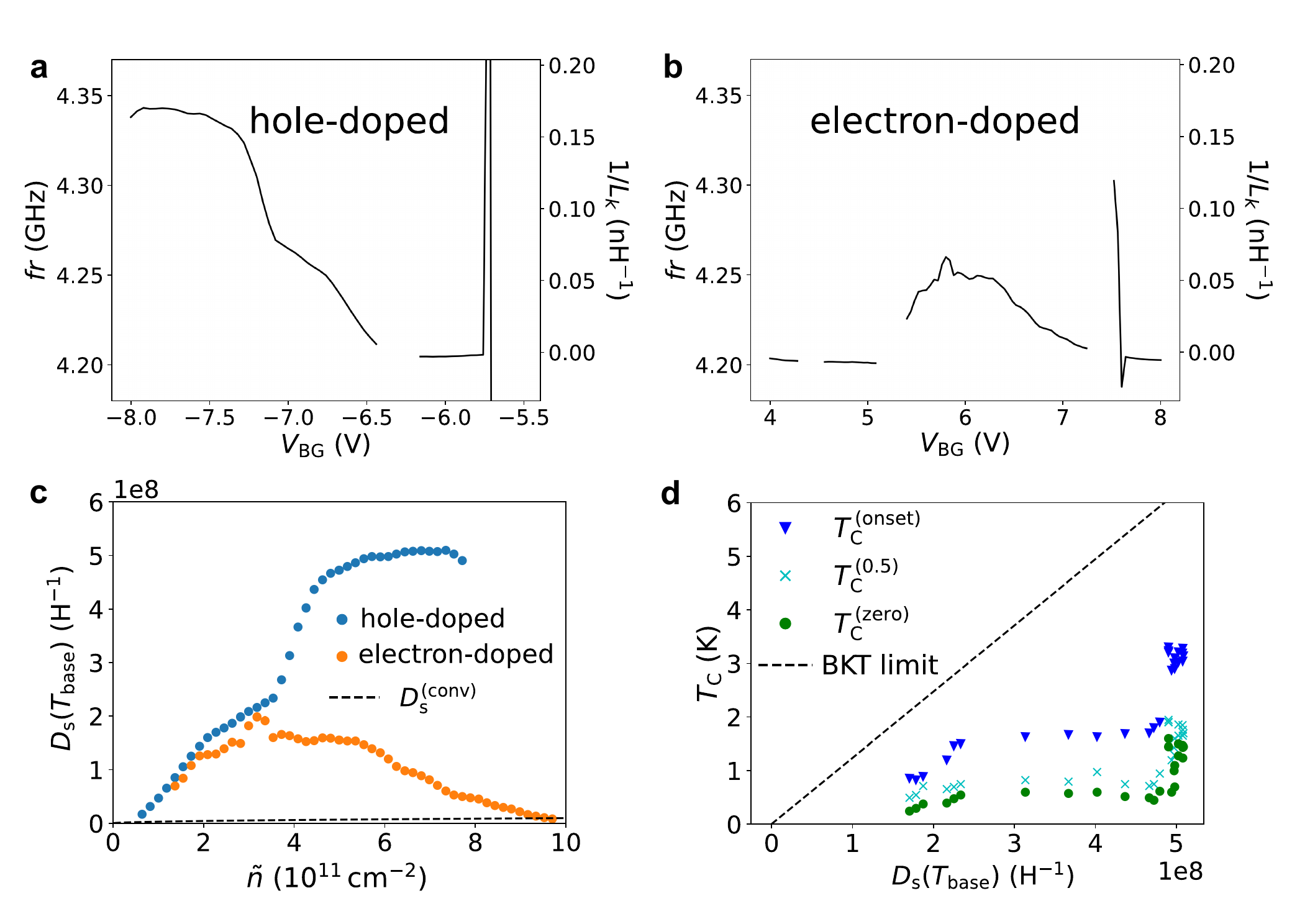}
    \caption{\textbf{Superfluid stiffness, carrier density, and critical temperature in the second device.} 
    \textbf{a, b,} Frequency shift and inverse of the kinetic inductance as a function of $V_{\mathrm{BG}}$ in hole-doped (a) and electron-doped (b) regimes. 
    \textbf{c,} Superfluid stiffness $D_{\mathrm{s}}$ at base temperature $T_{\mathrm{base}}$ as a function of effective carrier density $\tilde{n}$, measured with respect to $|\nu|=2$. 
    The black dashed curve estimates the conventional contribution to the superfluid stiffness from single-band Fermi liquid theory: $D_{\mathrm{s}}^{\mathrm{(conv)}}=e^{2}\tilde{n}v_{\mathrm{F}}/\hbar k_{\mathrm{F}}$. 
    \textbf{d,}  Critical temperature $T_{\mathrm{C}}$ 
    and corresponding superfluid stiffness $D_{\mathrm{s}}$ at base temperature $T_{\mathrm{base}}$ as tuned by $V_{\mathrm{BG}}$. The black dashed line represents the BKT limit $T_{\mathrm{C}}=\pi\hbar^{2}D_{\mathrm{s}}(T_{\mathrm{base}})/8e^{2}k_{\mathrm{B}}$.}
    \label{fig:B3ST5D}
\end{figure}

Fig.~\ref{fig:B3ST5D} presents the magnitude of the superfluid stiffness, which like the first device, is much larger than the conventional Fermi liquid theory would predict (Fig. \ref{fig:B3ST5D}c). The relation between $T_{\mathrm{C}}$ and $D_{\mathrm{s}}$ is consistent with the BKT theory (Fig. \ref{fig:B3ST5D}d).

\begin{figure}[H]
    \centering
    \includegraphics[width=1\textwidth]{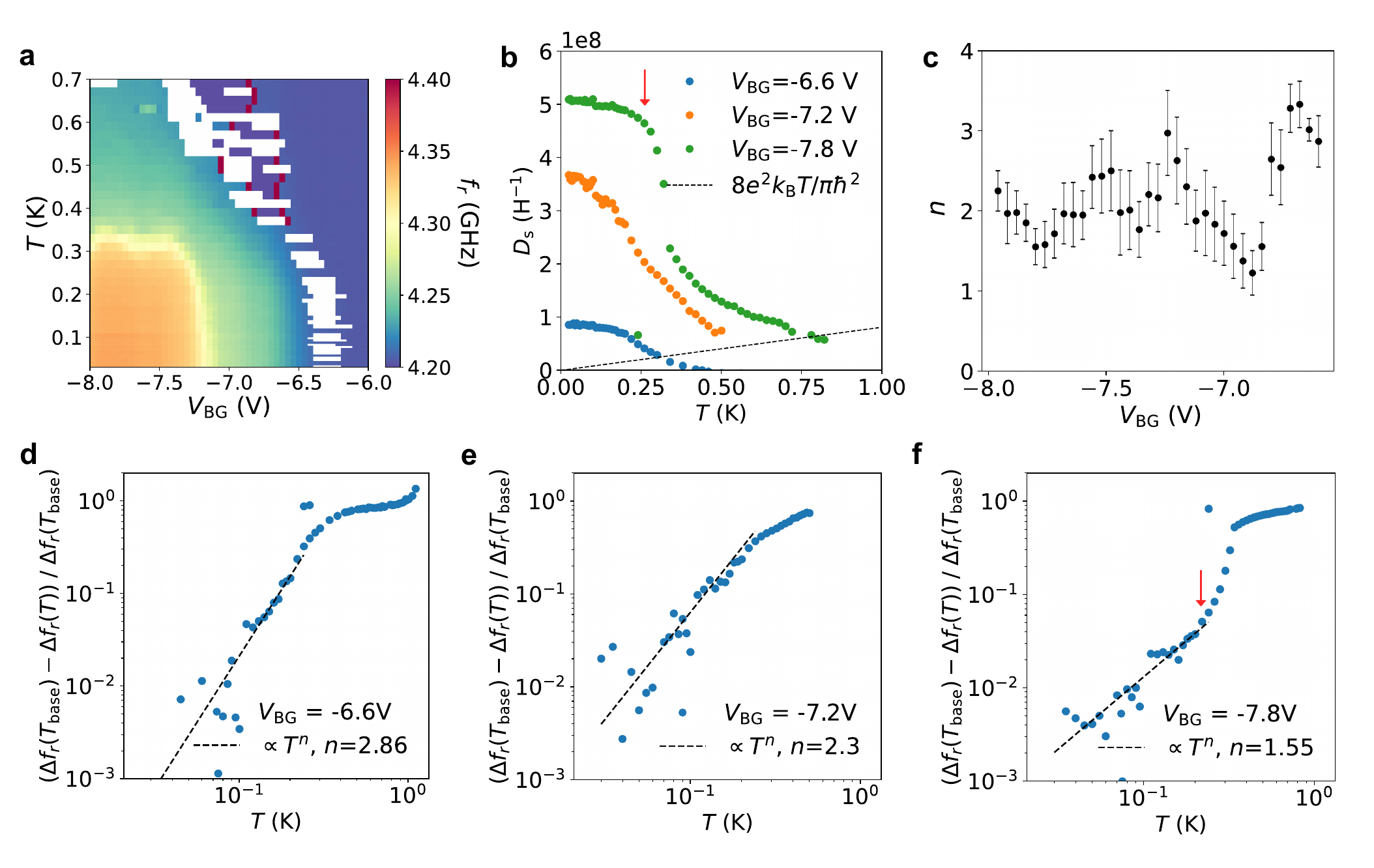}
    \caption{\textbf{Temperature dependence in the second device.} 
    \textbf{a,} Resonance frequency $f_{\mathrm{r}}$ as a function of temperature and $V_{\mathrm{BG}}$. 
    \textbf{b,} Temperature dependence of the measured $D_{\mathrm{s}}$ at $V_{\mathrm{BG}}$=-6.6, -7.2, and -7.8 V and $8e^{2}k_{\mathrm{B}}T/\pi \hbar^{2}$ (black dashed line), where $T_{\mathrm{BKT}}$ is determined by their intersection.
    \textbf{c,} $V_{\mathrm{BG}}$ dependence of the exponent determined from the power-law fitting at $T<0.3\;T_{\mathrm{C}}$. 
    \textbf{d, e, f} Power-law fitting of $(\Delta f_{\mathrm{r}}(T_{\mathrm{base}})-\Delta f_{\mathrm{r}}(T))\,/\,\Delta f_{\mathrm{r}}(T_{\mathrm{base}})$ at $V_{\mathrm{BG}}$=-6.6, -7.2, and -7.8 V.}
    \label{fig:B3ST5-T}
\end{figure}

Fig.~\ref{fig:B3ST5-T} shows the temperature dependence in the hole-doped SC region. 
Although the temperature dependence of $f_{\mathrm{r}}$ and $D_{\mathrm{s}}$ have multiple steps due to having multiple regions with different $T_{\mathrm{C}}$ values (Fig.~\ref{fig:B3ST5-T}a and b), the low-temperature power-law analysis is still a valid means to extract results in the low-excitation limit. 
Fig.~\ref{fig:B3ST5-T}c is the power-law exponent $n$ obtained from the fitting at $T<0.3\;T_{\mathrm{C}}$. The exponent is $n=1.3\sim3.2$, which is comparable with the first device in indicating an anisotropic paring.

\begin{figure}[H]
    \centering
    \includegraphics[width=1\textwidth]{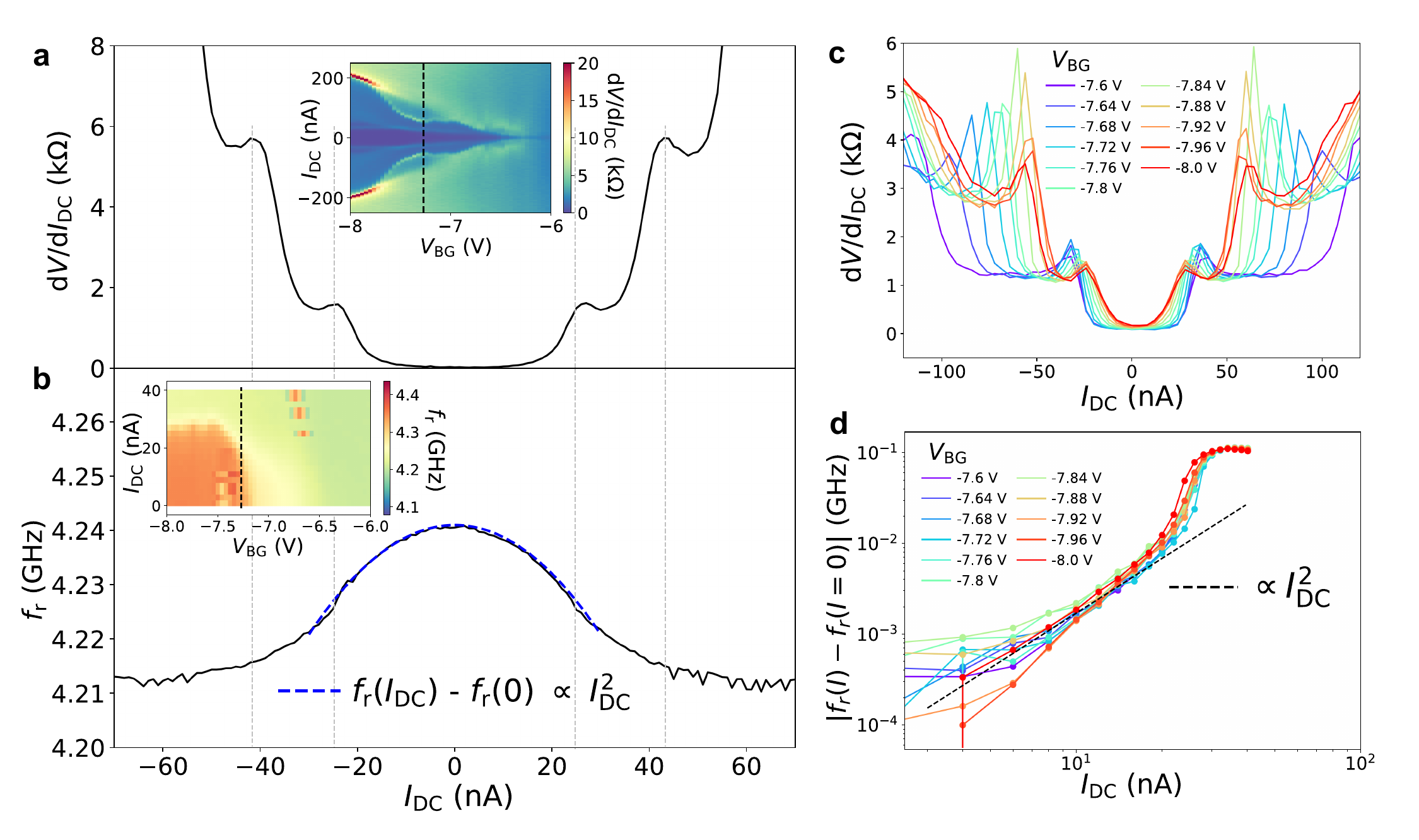}
    \caption{\textbf{DC bias current dependence in the second device.} 
    \textbf{a,} $I_{\mathrm{DC}}$ dependence of the DC resistance $\mathrm{d}V/\mathrm{d}I_{\mathrm{DC}}$ at $V_{\mathrm{BG}}$=-7.3 V. Inset depicts the DC resistance as a function of $I_{\mathrm{DC}}$ and $V_{\mathrm{BG}}$.
    \textbf{b,} $I_{\mathrm{DC}}$ dependence of $f_{\mathrm{r}}$ at $V_{\mathrm{BG}}$=-7.3 V. The blue dashed curve is the quadratic fit. Inset depicts $f_{\mathrm{r}}$ as a function of $I_{\mathrm{DC}}$ and $V_{\mathrm{BG}}$.
    \textbf{c,} $I_{\mathrm{DC}}$ dependence of the DC resistance over the hole-doped SC region.  
    \textbf{d,} $I_{\mathrm{DC}}$ dependence of $|f_{\mathrm{r}}(I)-f_{\mathrm{r}}(I=0)|$ over the hole-doped SC region using a logarithmic scale. The black dashed line indicates the quadratic dependence.}
    \label{fig:B3ST5-bias}
\end{figure}

The DC bias dependence below the smallest $I_{\mathrm{C}}$ showed quadratic behavior and is also consistent with the first device (Fig.\ref{fig:B3ST5-bias}). 

\begin{figure}[H]
    \centering
    \includegraphics[width=1\textwidth]{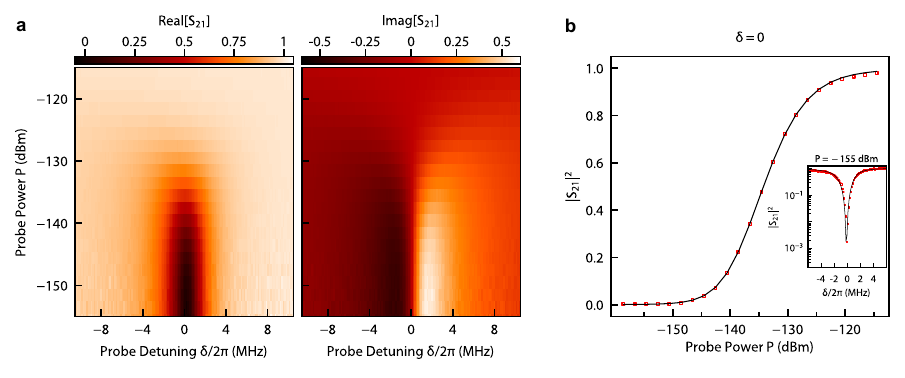}
    \caption{\textbf{Device temperature measurements with superconducting qubit spectroscopy.} 
    \textbf{a,} Real (left) and imaginary (right) components of the transmission spectrum of a coherent probe incident on a transmon qubit coupled to a coplanar waveguide as a function of the qubit-probe detuning $\delta/2\pi$ and the probe power $P$. 
    \textbf{b,} Transmittance $|S_{21}|^2$ as a function of probe power P at zero qubit-probe detuning ($\delta/2\pi = 0$). The measured data are plotted in red, and the theoretical fit is plotted in black. The inset shows the extinction in transmission in the form of a Lorentzian for a resonant probe at power $P$ = -155 dBm. The qubit acts as a single-photon mirror to resonant photons that arrive at the qubit on-average in the single-photon limit and smaller. We extract the qubit-waveguide coupling rate $\gamma/2\pi = 3.2$ MHz and the device temperature $T = 41.21 \pm 1.39$ mK. }
    \label{fig:spec}
\end{figure}

\section{Device Effective Temperature Measurement}
We measure the effective temperature of superconducting qubits that share the same mixing chamber of the dilution refrigerator and similar driving lines~\autocite{kannan2023demand}. We strongly couple a transmon qubit~\autocite{Koch2007} at frequency $\omega/2\pi = 4.88$ GHz to an open coplanar waveguide with coupling strength $\gamma /2 \pi = 3.2$ MHz. We send a coherent probe into the waveguide and measure the elastic scattering of the qubit~\autocite{kannan2023demand}.  

In the single-photon limit and less, the qubit absorbs and re-emits the photon in the forward and backward directions with a phase shift. The drive and re-emitted photons ideally perfectly interfere destructively in the forward direction and constructively in the reverse direction. This results in the extinction of the transmission signal for probes with low average photon numbers $|\alpha|^2\ll 1$ ~\autocite{Hoi2011, Astafiev2010, Hoi2013}. The master equation for the simplified model of a single qubit coupled to a waveguide is given by~\cite{Mirhosseini2019},
\begin{equation}
    \partial_t \hat{\rho} = \frac{1}{i\hbar}\big[\hat H,\hat\rho\big] + (\bar n_\mathrm{th} + 1)\gamma D\big[\hat{\sigma}^-\big]\hat\rho + \bar n_\mathrm{th}\gamma D\big[\hat{\sigma}^+\big]\hat\rho + \frac{\gamma_\phi}{2} D\big[\hat{\sigma}_z\big]\hat\rho.
\end{equation}
Here the single-qubit Hamiltonian is $\hat H = \frac{1}{2}\delta \hat\sigma_z  + \frac{1}{2}\Omega_p\hat\sigma_x$, where $\gamma_{\phi}$ is the pure dephasing rate of the qubit, $\delta = \omega - \omega_p$ is the qubit-probe detuning, the Lindblad dissipator is $D[\hat{O}] = \hat{O}\hat\rho\hat{O}^\dagger - \frac{1}{2} \{\hat{O}^\dagger\hat{O},\hat\rho\}$, and $\Omega_p = \sqrt{2\gamma P/\hbar\omega_\mathrm{p}}$ is the drive strength of the probe with power $P$. The qubit is coupled to a thermal bath of photons present in the waveguide, with temperature $T$ and average thermal photon number $\bar n_\mathrm{th} = 1/(e^{\frac{\hbar\omega_\mathrm{q}}{k_\mathrm{B}T}} - 1)$.The operators $\hat{\sigma}^+$ and $\hat{\sigma}^-$ raise and lower the qubit state, respectively.  

Assuming that the probe propagates towards the right, the rightward propagating output of the waveguide can be determined via input-output theory:
\begin{equation}
    \hat a_\mathrm{R} = \hat a_\mathrm{R}^\textrm{in} + \sqrt{\frac{\gamma}{2}}\hat \sigma^{-}.
\end{equation}
Therefore, we can calculate $S_{21} = \langle\hat a_\mathrm{R}\rangle/\langle\hat a_\mathrm{R}^\textrm{in}\rangle$~\autocite{Mirhosseini2019},
\begin{equation}
    S_{21}(\delta, \Omega_p) = 1 - \frac{\gamma(1 + i\delta/\gamma_2^\mathrm{th})}{2 \gamma_2^\mathrm{th}(2\bar n_\mathrm{th} + 1)[1 + (\delta/\gamma_2^\mathrm{th})^2 +  \Omega_\mathrm{p}^2/(\gamma_1^\mathrm{th}\gamma_2^\mathrm{th})]}.
\end{equation}
We define a thermally enhanced decay and dephasing rate, $\gamma_1^\mathrm{th} = (2\bar n_\mathrm{th} + 1)\gamma$ and $\gamma_2^\mathrm{th} = \gamma_1^\mathrm{th}/2 + \gamma_\phi$, where $\gamma_2 = \gamma/2 + \gamma_\phi$ is the bare decoherence rate. Transmission measurements as a function of probe power $P$ and detuning $\delta$, as shown shown in Fig.~\ref{fig:spec}, enable us to extract the device effective temperature $T = 41.21 \pm 1.39$ mK. These measurements also calibrate the absolute power of microwave tones incident on devices.

\end{document}